\documentclass[12pt]{article}

\usepackage[english]{babel}
\usepackage{amsmath}
\usepackage{amsfonts}
\usepackage{amssymb}
\usepackage{a4wide}
\usepackage{graphicx}
\usepackage{cite}
\usepackage{color}
\usepackage[dvipsnames]{xcolor}
\usepackage [autostyle, english = american]{csquotes}
\MakeOuterQuote{"}
\usepackage[colorlinks]{hyperref}
\hypersetup{pageanchor=false,linkcolor=NavyBlue,citecolor=NavyBlue,urlcolor=RoyalPurple}

\usepackage{slashed}
\usepackage{float}
\usepackage{dsfont}
\usepackage{yfonts}
\usepackage{appendix}
\usepackage{xcolor}
\usepackage{mathtools}
\DeclareMathOperator*{\sign}{\textrm{sign}}

\allowdisplaybreaks[1]

\newcommand{\C}{\mathbb{C}}
\newcommand{\p}{\partial}

\newcommand{\ve}{\varepsilon}
\newcommand{\der}{\textrm{d}}
\newcommand{\Ld}{\mathcal{L}}
\newcommand{\Lo}{\mathcal{O}}
\newcommand{\tr}{\textrm{tr}}
\newcommand{\A}{\mathcal{A}}
\newcommand{\un}{\mathds{1}}
\newcommand{\ket}[1]{| #1 \rangle}
\newcommand{\bra}[1]{\langle #1 |}

\newcommand{\ch}{\mathfrak{h}}
\newcommand{\cA}{\textgoth{A}}
\newcommand{\SU}{\textrm{SU}}

\definecolor{navy}{rgb}{0.0,0.0,0.5}

\newcounter{MBQ}

\newcounter{ASQ}

\newcounter{PHQ}

\numberwithin{equation}{section}

\begin{document}
\thispagestyle{empty}
\allowdisplaybreaks

\begin{flushright}
{\small
TUM-HEP-1347/21\\
2106.09054 [hep-th]\\
February 11, 2022}
\end{flushright}

\vskip1.5cm
\begin{center}
{\Large\bf\boldmath 
Double copy for Lagrangians at trilinear order}
\vspace{\baselineskip}

\vspace{1cm}
{\sc M.~Beneke}${}^{a}$,  
{\sc P.~Hager}${}^{a}$
and 
{\sc A. F. Sanfilippo}${}^{a}$\\[5mm]
${}^a${\it Physik Department T31,\\ 
Technische Universit\"at M\"unchen,\\ 
James-Franck-Stra\ss e~1,\\ 
D - 85748 Garching, Germany
}\\[0.5cm]

\vspace*{1cm}
\textbf{Abstract}\\
\vspace{1\baselineskip}
\parbox{0.9\textwidth}{
We present a novel double-copy prescription for gauge 
fields at the Lagrangian level and apply it to the original 
double copy, couplings to matter and the soft theorem. The 
Yang-Mills Lagrangian in light-cone gauge is mapped directly to the $\mathcal{N}=0$ supergravity Lagrangian in light-cone gauge to trilinear order, and we show that the obtained result is manifestly equivalent to Einstein gravity at tree level up to this order. The application of the double-copy prescription to couplings to matter is exemplified by scalar and fermionic QCD and finally the soft-collinear effective QCD Lagrangian. The mapping of the latter yields an effective description of an energetic Dirac fermion coupled to the graviton, Kalb-Ramond, and dilaton fields, from which the fermionic gravitational soft and next-to-soft theorems follow.}
\end{center}

\newpage
\setcounter{page}{1}


\newpage

\section{Introduction}

The double-copy property refers to 
the observation that tree-level graviton scattering amplitudes in 
Einstein-Hilbert gravity can be obtained by ``squaring'' tree-level 
Yang-Mills scattering amplitudes with the replacement of colour factors by kinematic factors.

The first indication of the connection between gauge theory and gravity was provided by the Kawai-Lewellen-Tye (KLT) relations \cite{KLT}. The modern formulation of the link between gauge-theory and gravity amplitudes was then conjectured by Bern, Carrasco and Johansson in \cite{BCJconjecture} and then proven for tree-level amplitudes in \cite{BCJ}, thereafter dubbed BCJ double copy or just double copy. The proof rests on the conjecture that the kinematic numerators of a tree-level Yang-Mills amplitude can always be brought to a form in which they satisfy the same algebraic properties as the colour factors. This conjecture is referred to as colour-kinematics duality. Using this property the validity of the double-copy formula for tree-level amplitudes  can be obtained using BCFW recursion relations \cite{BCFW1,BCFW2}. The immediate extension to loop amplitudes via generalized unitarity was also discussed in \cite{BCJ}. This formulation of the duality has since then seen a plethora of further extensions, see \cite{DC} for a comprehensive review. The property also often generalizes to the couplings to matter \cite{Chiodaroli:2013upa,Johansson:2014zca,QCDproof,DCmatter} and to (supersymmetric) effective field theories, such as the non-linear sigma model and Dirac Born-Infeld theory, see \cite{Chen:2013fya,scatt,Du:2016tbc,Cheung:2016prv,chiod} and references therein. 
The double-copy property is usually thought to hold at the 
level of on-shell amplitudes, not at the level of Lagrangians, 
reflecting the fact that the Lagrangian carries redundant 
information if one is only interested in scattering amplitudes. 
Its deeper origin and scope are not fully understood.

The present work is motivated by the attempt to construct 
the soft theorem for gravitons from the soft theorem for gluons via the double-copy of soft-collinear Lagrangians.\footnote{The double-copy of soft emission was already studied 
for the particular case of a $2\to 2$ scattering 
process at amplitude level 
in \cite{SabioVera:2014mkb}.} Such an approach 
can be seen as  complementary to the elegant derivation 
within the spinor-helicity formalism \cite{softQCD,CS}, where 
the double-copy property of 
tree amplitudes follows from the corresponding property of 
the three-point amplitude and the recursive construction of 
$n$-point amplitudes \cite{BCJ}. The soft-collinear 
effective (SCET) Lagrangian 
employs only fields that correspond directly to 
the collinear and soft on-shell degrees of freedom in a hard 
scattering process. For example, the gauge-invariant collinear 
gluon field $g_s \mathcal{A}^\mu = W^\dagger [i D^\mu W]$, where 
$W$ denotes a semi-infinite Wilson line along the light-like 
direction $n_+^\mu$, satisfies $n_+\cdot \mathcal{A} = 0$, and 
the Lagrangian can be expressed entirely in terms of the 
two transverse components of $\mathcal{A}^\mu$. 
In the absence of soft fields, 
the collinear Lagrangian of SCET provides a gauge-invariant 
formulation of the theory in light-cone gauge 
\cite{Beneke:2002ni,Beneke:2002ph}. While 
the soft-collinear Lagrangian does contain (irrelevant) off-shell 
information, its formulation in terms of physical degree-of-freedom 
fields suggests that the double-copy property could be 
more easily made manifest in this framework.

In this paper we show that indeed a novel mapping can be found which parallels colour-kinematics duality as applied to tree-level amplitude numerators but which acts directly on fields and operations on fields and is therefore not limited to the application to pure gauge theory. We first use it to map the Lagrangian of Yang-Mills theory to the one of $\mathcal{N}=0$ supergravity up to cubic order and show its equivalence to Einstein gravity for tree-level amplitudes with only gravitons in the initial and final states. 
We find that this becomes possible once light-cone gauge has 
been fixed, in agreement with the expectation that 
gauge redundancy obscures the link between the two theories. The mapping also applies to couplings to matter, once light-cone gauge has been fixed for the gauge field. We exemplify this first by the scalar and fermionic QCD Lagrangians to trilinear order, on which the mapping produces the correct 
couplings of gravitons at the same order. We then go on to show that the soft and next-to-soft theorems for the 
emission of a gluon and graviton from matter can be similarly 
related by applying the same mapping to the soft-collinear interaction 
terms directly in the SCET Lagrangian. 

In the present treatment, the prescription to obtain the gravitational 
theory from the gauge theory can be interpreted as trading the gauge
 group of Yang-Mills theory, $\SU(N)$, for the gauge group of 
Einstein gravity, $\textrm{Diff}(M)$. This effectively amounts to the substitution given in \eqref{finalvector} below together with a pairing prescription, offering a geometric intuition of colour-kinematics duality. As a consequence of the 
manifest duality between the Lagrangians, the three-point vertex 
Feynman rules in light-cone gauge satisfy an explicit squaring
relation, and they do so off-shell. 

Earlier work towards a geometric understanding of colour-kinematics duality can be found in \cite{Monteiro:2011pc,Fu:2016plh}. In \cite{Monteiro:2011pc} the authors restricted themselves to the self-dual sector of Yang-Mills theory and there they identified the Poisson algebra as the kinematic analogue of the colour algebra. The double-copy link between the self-dual sectors of gauge and gravitational theories was further studied in \cite{Campiglia:2021srh}. In \cite{Fu:2016plh} the authors considered the full Yang-Mills theory and highlighted the link between the structure of the kinematic pieces of the Yang-Mills vertices and the Lie algebra of vector fields. Specifically, they showed that up to four-point amplitudes the origin of the duality between colour and kinematic numerators can be explained by identifying the Drinfeld double of the Lie algebra of vector fields as the underlying kinematic algebra. 

Previous attempts towards a Lagrangian version of the double copy started in \cite{EHfromQCD}, where tree-level QCD amplitudes were double-copied to gravity amplitudes up to five points. A Lagrangian which would reproduce the obtained gravitational amplitudes was written down, and it was argued that it is equivalent to the Einstein-Hilbert Lagrangian for on-shell amplitudes. However, this is not what we have in mind when referring to the double copy of Lagrangians, since we would like to directly map Lagrangians onto Lagrangians without having to take a detour through the amplitudes.

In \cite{KLTDC} it was shown that the Einstein-Hilbert Lagrangian up to quartic order could be rewritten adopting light-cone gauge, expressing it explicitly in terms of the two propagating helicity modes of the graviton, and then performing a canonical field redefinition to obtain a form where the coefficients of the Fourier-transformed Lagrangian follow from applying the KLT relations to the Fourier-transformed coefficients of the Cachazo-Svrček-Witten Lagrangian \cite{CSW}. This approach still relies on the knowledge of colour-kinematics duality at the amplitude level and, further, on a non-standard representation of both the Einstein-Hilbert and Yang-Mills Lagrangians, which obscures whether the results also apply to scattering processes which cannot be obtained via the maximally-helicity-violating rules, such as the $(++-)$ helicity amplitude.

In \cite{BCJ} a modified but equivalent version of the standard Yang-Mills Lagrangian, yielding Feynman rules already in BCJ dual form up to quintic order, was found and brought to a form containing only trilinear interaction terms using auxiliary fields. The obtained expression was then stripped of colour factors and "squared" order by order, meaning that two copies of the colour-stripped expression were multiplied together, and the pairs of gauge fields were identified with graviton fields, to obtain a Lagrangian which leads to the correct double-copy amplitudes. This squaring procedure was corroborated by the observations in\cite{Hohm:2011dz} that the $\mathcal{N}=0$ supergravity Lagrangian can be written in an "index-factorized" form, which means that the index structure allows for two distinct sets of indices to be defined, which never get contracted with each other. This statement was refined in \cite{factorizedEH,Cheung:2017kzx}, where it was shown that this holds as well for the Einstein-Hilbert Lagrangian alone, i.e. without the dilaton and 2-form. The same procedure was also found to apply to the action of the non-linear sigma model to obtain the one for the special Galileon in \cite{Cheung:2016prv}. Again, in \cite{BCJ} the Yang-Mills Lagrangian was written in an unusual way to get a gravitational Lagrangian in terms of an infinite tower of auxiliary fields, whereas we would like to find a direct mapping from the Yang-Mills Lagrangian, expressed only in terms of the standard gauge field, to the Einstein-Hilbert Lagrangian, expressed in terms of the metric perturbation. The approach in \cite{BCJ} was adapted to the Yang-Mills Lagrangian in light-cone gauge in \cite{Vaman:2014iwa}, where the explicit quintic order modification to the Lagrangian was given, and finally extended in \cite{Tolotti:2013caa}, where an algorithm to construct an effective Yang-Mills Lagrangian generating Feynman rules which lead to arbitrary $n$-point amplitudes with kinematic numerators already in BCJ form was found. Both \cite{Vaman:2014iwa,Tolotti:2013caa} did not explore how to obtain a gravity Lagrangian from their result. For another approach towards constructing the trilinear gravity Lagrangian, see \cite{cubicck}.

More recently, in \cite{BRSTDC,Borsten:2021hua} it was shown abstractly that an equivalent representation of the full $\mathcal{N}=0$ supergravity Lagrangian can be obtained by squaring a redefined version of the BRST Yang-Mills Lagrangian, containing only cubic interaction terms. This construction was extended in \cite{Borsten:2021rmh} to obtain actions which generate Feynman rules which make color-kinematics duality manifest also for loop diagrams. Finally, in \cite{EHcubic} the BRST Einstein-Hilbert Lagrangian up to cubic order was explicitly obtained by "squaring" the trilinear BRST Yang-Mills Lagrangian, and the ghost fields proved crucial for the success of their approach. This construction was recently extended to apply to generic homogeneous spaces in \cite{Borsten:2021zir}. The result in \cite{EHcubic} bears the closest resemblance to one of our main results. However, the approach is different, since we will fix the ghost-free light-cone gauge and only work with physical degrees of freedom. While the idea of fixing light-cone gauge to eliminate redundancies at the Lagrangian level is not new, to the best of our knowledge, it has never been used in the context of a direct map from the Yang-Mills to the Einstein-Hilbert Lagrangian, nor were couplings to matter investigated using this method. 


\section{Colour-kinematics duality prescription for fields}
\label{rules}

The main new idea underlying the following mappings is to replace the Lie algebra of the gauge group $\SU(N)$ of Yang-Mills theory by the algebra of tangent vector fields on the spacetime manifold $M$, which are the generators of diffeomorphisms on $M$.
In the following, we will assume a flat background metric $\eta_{\mu\nu}
$ with signature $(+,-,-,-)$. 

Specifically, we map the colour generators $T^a$ to colour-stripped gauge tangent vector fields times a factor of $-i$,\footnote{The 
tilde on the $A^\mu$ field is temporarily introduced to distinguish it from the $A^\mu$ field that multiplies the generator in the equation below.}
\begin{equation}
T^a\mapsto-i\tilde{A}^{\mu}(x)\p_{\mu}\,.
\label{vector}
\end{equation}
To define the classical double-copy field $H_{\mu\nu}$, we 
replace the sum over the adjoint colour index $a$ by a 
convolution 
\begin{equation}
A^a_{\mu}(x)T^a\mapsto-i(A_{\mu}*\tilde A_{\nu})(x)\p^{\nu}\equiv -i I\big[A_{\mu}\big|\tilde A_{\nu}\big]\p^{\nu}\equiv -iH_{\mu\nu}(x)\p^{\nu}\,,
\label{finalform}
\end{equation}
in analogy to \cite{BCJ}. We follow the notation of \cite{fatgrav}, where the double-copy field is 
called "fat graviton". We introduced the pairing operation $I$, and, in a product of multiple fields on which we apply this mapping below, we introduce labels to clarify which fields belong together. 

The gauge field $A^a_{\mu}(x)T^a$ is a Hermitian matrix with respect to the inner product on $\C^N$, so we need to map it to an object which is still Hermitian, now with respect to 
the $L^2$ inner product
\begin{equation}
\langle\phi,\psi\rangle\equiv\int\der^4x\;\phi^*(x)\psi(x)\,.
\label{innerprod}
\end{equation}
The derivative $-i\p_{\mu}$ is Hermitian with respect to \eqref{innerprod}, but the object $-iH_{\mu\nu}(x)\p^{\nu}$ is not, since, in general, $\p^{\nu}H_{\mu\nu}$ is non-vanishing. To remedy this, we need to work with the double-copy field in a transverse gauge, 
such that it satisfies
\begin{equation}
\p^{\nu}H_{\mu\nu}(x)=0\,.
\end{equation}
To ensure this, we construct it from $A^a_{\mu}$ in a gauge which satisfies 
\begin{equation}
\p^{\mu}A^a_{\mu}(x)=0+\Lo(A^2)\,,
\end{equation}
since we are defining a linear prescription. The mapping \eqref{finalform} is then equivalent to the manifestly Hermitian mapping
\begin{equation}
\begin{aligned}
T^a& \mapsto-\frac{i}{2}\,[\tilde{A}^{\mu}(x)\overset{\rightarrow}{\p_{\mu}}-\overset{\leftarrow}{\p_{\mu}}\tilde{A}^{\mu}(x)]\,,\label{finalvector}\\
A^a_{\mu}(x)T^a &\mapsto-\frac{i}{2}\,[H_{\mu\nu}(x)\overset{\rightarrow}{\p^{\nu}}-\overset{\leftarrow}{\p^{\nu}}H_{\mu\nu}(x)]\,,
\end{aligned}
\end{equation}
when appearing under an integral, as in \eqref{innerprod}.

The partial derivatives $\p_{\mu}$ are the basis vectors of the tangent space at each point of Minkowski space $M$. We denote the set of tangent vector fields on $M$ as $\Gamma(TM)$ and the ring of real-valued functions on $M$ as $\mathcal{F}(M)$. We use the standard definitions of the Minkowski metric tensor $\eta$, 
\begin{equation}
\begin{aligned}
\eta&\colon \Gamma(TM)\times \Gamma(TM) \to \mathcal{F}(M)\,,\\[0.1cm]
\eta(V(x),W(x)) &= V^{\mu}(x)W^{\nu}(x)\eta(\p_{\mu},\p_{\nu})\equiv V^{\mu}(x)W^{\nu}(x)\eta_{\mu\nu}\,,\label{eta}
\end{aligned}
\end{equation}
as well as of the Lie bracket of vector fields in $\Gamma(TM)$,\footnote{In the following, derivatives (and inverse derivatives) act only on the object immediately next to them, except if multiple objects are grouped in a bracket, such as $\p_{\mu}(fg)$.}
\begin{equation}
\begin{aligned}
[\cdot,\cdot]_L \colon \Gamma(TM)&\times \Gamma(TM)\rightarrow\Gamma(TM)\,,\\[0.1cm]
[V(x),W(x)]_L =&(V^{\mu}\p_{\mu}W^{\nu}-W^{\mu}\p_{\mu}V^{\nu})\p_{\nu}\,.\label{lie}
\end{aligned}
\end{equation}
 
Armed with these definitions, we are now ready to define the remaining mappings of colour-kinematics duality. The commutator $[\cdot,\cdot]$ of colour generators, which satisfies
\begin{equation}
[T^a,T^b]=if^{abc}T^c
\label{liea}
\end{equation}
with the totally antisymmetric structure constants $f^{abc}$, is mapped to the Lie bracket of tangent vectors $[\cdot,\cdot]_L$ times a factor of $-1/2$,\footnote{The factor $-1/2$ is related to the fact that the standard double-copy refers to the  rescaled 
structure constants defined in \eqref{eq:ftilde} below.}
\begin{equation}
[\cdot,\cdot]\mapsto-\frac{1}{2}\,[\cdot,\cdot]_L\,,
\end{equation}
such that
\begin{equation}
[T^a,T^b]\stackrel{\eqref{vector}}{\mapsto}-\frac{1}{2}[-iA^{1\mu}\p_{\mu},-iA^{2\nu}\p_{\nu}]_L=\frac{1}{2}(A^{1\mu}\p_{\mu}A^{2\nu}-A^{2\mu}\p_{\mu}A^{1\nu})\p_{\nu}\,.
\end{equation}
The superscripts 1, 2 refer to the labels mentioned above.

The operation of taking the trace over colour generators is mapped to the operation of taking the inner product of tangent vectors by means of the metric tensor $\eta$ times a factor $1/2$, which accounts for the normalization of the colour generators
\begin{equation}
\tr(T^aT^b)=\frac{1}{2}\,\delta^{ab}\,.
\end{equation}
We have to ensure that the symmetries of the colour indices of the generators before the mapping and the labels of the colour-stripped fields after the mapping match. If they do not, we sum over all necessary permutations of the labels until they do. Therefore, we have:
\begin{equation}
\tr(\cdot\cdot)\mapsto\frac{1}{2}\sum_{\textrm{perm.}}\eta(\cdot,\cdot)\,.
\label{tracemap}
\end{equation}
Single generators are traceless, therefore 
\begin{equation}
0=\tr(T^a\un)\mapsto\frac{1}{2}A^{\mu}\eta(-i\p_{\mu},0)=0\,,
\end{equation}
which means that the neutral element of the algebra of $\SU(N)$ is mapped to the trivial tangent vector field 0. In the case of only two generators, we simply have
\begin{equation}
\tr(T^aT^b)\mapsto\frac{1}{2}\eta(-iA^{1\mu}\p_{\mu},-iA^{2\nu}\p_{\nu})=-\frac{1}{2}A^{1\mu}A^{2\nu}\eta(\p_{\mu},\p_{\nu})=-\frac{1}{2}A^1_{\mu}A^{2\mu}\,.
\end{equation}
Finally, a totally antisymmetric object, such as $\tr(T^a[T^b,T^c])$, gets mapped to
\begin{align}
\begin{split}
\tr(T^a[T^b,T^c]) &\mapsto-\frac{i}{4}\bigg[\eta\Big(A^{1\mu}\p_{\mu},[A^{2\nu}\p_{\nu},A^{3\rho}\p_{\rho}]_L\Big)+\eta\Big(A^{2\mu}\p_{\mu},[A^{3\nu}\p_{\nu},A^{1\rho}\p_{\rho}]_L\Big)\\
&\quad\:\,\phantom{-\frac{i}{4}\bigg[}
+\eta\Big(A^{3\mu}\p_{\mu},[A^{1\nu}\p_{\nu},A^{2\rho}\p_{\rho}]_L\Big)\bigg]
\end{split}\nonumber\\
\begin{split}\label{interf1}
&=-\frac{i}{4}\bigg[A^{1\mu}(A^{2\nu}\p_{\nu}A^3_{\mu}-A^{3\nu}\p_{\nu}A^2_{\mu})+A^{2\mu}(A^{3\nu}\p_{\nu}A^1_{\mu}-A^{1\nu}\p_{\nu}A^3_{\mu})\\
&\quad\:\,\phantom{-\frac{i}{4}\bigg[}+A^{3\mu}(A^{1\nu}\p_{\nu}A^2_{\mu}-A^{2\nu}\p_{\nu}A^1_{\mu})\bigg]\,,
\end{split}
\end{align}
which is totally antisymmetric in the labels. This is the reason why we introduced the sum over permutations of the labels in \eqref{tracemap}. Notice that the right-hand-side of \eqref{interf1} exactly coincides with the totally antisymmetrized kinematic part of the Yang-Mills trilinear Lagrangian before gauge-fixing. The fact that the kinematic part of the three-point Yang-Mills vertex Feynman rule, which arises from the totally antisymmetrized expression \eqref{interf1}, can be rewritten in terms of Lie brackets and the Minkowski metric tensor was noticed in \cite{Fu:2016plh}.\footnote{A connection between the structure constants and the Lie bracket is also established in \cite{Cheung:2016prv,Cheung:2020djz}, although different theories and replacement prescriptions are considered than the present ones.}
Lastly, we have to trade the strong coupling $g_s$ for the gravitational one,
\begin{equation}
g_s\mapsto\frac{\kappa}{2}\,,
\end{equation}
where $\kappa\equiv\sqrt{32\pi G_N}$. 

We have to map multiple products of gauge fields with their respective generators. As mentioned above, the operation $I$ pairs up multiple colour-stripped gauge fields according to their labels and identifies each pair with a double-copy field. It should also yield a vanishing result if we try to pair up two strings of fields of different length. We define it as
\begin{equation}
I\Big[A_{\mu}^1A_{\rho}^2...A_{\alpha}^n\Big|\tilde A_{\nu}^1\tilde A_{\sigma}^2...\tilde A_{\beta}^m\Big]\equiv \delta_{nm}H_{\mu\nu}H_{\rho\sigma}\cdot...\cdot H_{\alpha\beta}\,.
\label{pair1}
\end{equation}
Derivative or inverse derivative operators acting on single objects act cumulatively on the pairing of them, for instance
\begin{equation}
I\bigg[A^1_i\p_+A^{2,i}\frac{\p_j}{\p_+}A^{3,j}\bigg|\tilde A^1_k\p_+\tilde A^{2,k}\frac{\p_l}{\p_+}\tilde A^{3,l}\bigg]=H_{ik}\p^2_+H^{ik}\frac{\p_j\p_l}{\p^2_+}H^{jl}\,.
\label{pair2}
\end{equation}

Finally, we need to determine which types of particles are described by the double-copy field. To build $H_{\mu\nu}$, 
we take the convolution of two distinct copies of colour-stripped gauge fields, which contain two on-shell degrees of freedom each. Hence, the double-copy field $H_{\mu\nu}$ contains four degrees of freedom and has no particular symmetry in its indices. We can therefore decompose it as a sum of a symmetric-traceless part $h_{\mu\nu}$, an antisymmetric part $B_{\mu\nu}$, and a trace part $\phi$, in analogy to \cite{fatgrav},
\begin{equation}
H_{\mu\nu}(x)=h_{\mu\nu}(x)+B_{\mu\nu}(x)+C_{\mu\nu}\phi(x)\,,
\end{equation}
where
\begin{equation}
\begin{aligned}
h_{\mu\nu}(x)&\equiv\frac{1}{2}H_{(\mu\nu)}(x)-C_{\mu\nu}H^{\rho}_{\phantom{\mu}\rho}(x)\,,\\
B_{\mu\nu}(x)&\equiv\frac{1}{2}H_{[\mu\nu]}(x)\,,\\
\phi(x)&\equiv H^\mu_{\phantom{\mu}\mu}(x)\,.
\end{aligned}
\end{equation}
Here, $C_{\mu\nu}$ is a symmetric operator which satisfies
\begin{equation}
C^{\mu}_{\;\mu}\phi(x)=\phi(x)\,,
\end{equation}
and we used a shorthand notation for (anti-)symmetrization in a pair of indices:
\begin{equation}
x^{(\mu}y^{\nu)}\equiv x^{\mu}y^{\nu}+x^{\nu}y^{\mu}\,,\quad x^{[\mu}y^{\nu]}\equiv x^{\mu}y^{\nu}-x^{\nu}y^{\mu}\,.
\end{equation}
The symmetric-traceless part of $H_{\mu\nu}$ is identified with the graviton, the antisymmetric part with the Kalb-Ramond, and the trace part with the dilaton field; the theory describing the dynamics of these fields is usually referred to as $\mathcal{N}=0$ supergravity.


\section{Yang-Mills and Einstein-Hilbert Lagrangians in 
light-cone gauge}

\subsection{Yang-Mills}

We want to rearrange the Yang-Mills Lagrangian by using the gluon field in light-cone gauge, $\cA^a_{\mu}$, which is defined by
\begin{equation}
n_+\cdot\cA^a\equiv0
\end{equation}
with respect to a light-like vector $n^{\mu}_+$. Defining a second light-like vector $n^{\mu}_-$, which satisfies
$n_+\cdot n_-=2$, 
we can decompose the gluon field in light-cone gauge as
\begin{equation}
\cA^a_{\mu}=n_-\cdot\cA^a\,\frac{n_{+\mu}}{2}+\cA^a_{\mu_{\perp}}\,.
\label{lightconedec}
\end{equation}
The field $\cA^a_{\mu_{\perp}}$ denotes the components perpendicular to both $n^{\mu}_{\pm}$. From now on, we abbreviate $n_{\pm}\cdot v\equiv v_{\pm}$ for any vector $v^{\mu}$ and use Latin letters from the middle of the alphabet to denote transverse vector indices.

Only the transverse field components $\cA^a_i$ are physical and propagating, so we integrate out the field component $\cA^a_-$ by means of its equation of motion, which reads 
\begin{equation}
\cA^a_-=-\frac{2}{\p_+}\p_i\cA^{a,i}+2g_sf^{abc}\frac{1}{\p^2_+}(\cA^b_i\p_+\cA^{c,i})\,.
\label{a-}
\end{equation}
Indeed,
\begin{equation}
\p^{\mu}\cA^a_{\mu}=0+\Lo(\cA^2)\,,
\end{equation}
as required in Section \ref{rules}. The transverse indices are still lowered by means of the transverse metric $\eta_{ij}=-\delta_{ij}$ so that $\cA_i=-\cA^i$. 

Since the colour-kinematics duality mapping acts on the colour generators themselves, we make them explicit. The Yang-Mills Lagrangian in light-cone gauge then reads
\begin{align}
\Ld^{\textrm{kin}}_{\textrm{YM}}&=-\tr(T^aT^b)\,\p_{\mu}\cA^a_i\p^{\mu}\cA^{b,i}\,,\label{YMkin}\\
\Ld^{(g_s)}_{\textrm{YM}} &=\,2ig_s\tr(T^a[T^b,T^c])\bigg[\cA^a_i\p_+\cA^{b,i}\frac{\p_j}{\p_+}\cA^{c,j}+\cA^a_i\cA^b_j\p^j\cA^{c,i}\bigg]\label{tritr}\,,\\
\Ld^{(g_s^2)}_{\textrm{YM}}&=\,\frac{g^2_s}{2}\,\tr(T^b[T^c,[T^d,T^e]])\bigg[-2\cA^b_i\p_+\cA^{c,i}\frac{1}{\p^2_+}\bigg(\cA^d_j\p_+\cA^{e,j}\bigg)+\cA^b_i\cA^c_j\cA^{d,i}\cA^{e,j}\bigg]\,,\label{quadritr}
\end{align}
which is equivalent to \cite{lightcone}. 
From \eqref{YMkin}, we deduce the gluon propagator in momentum space,
\begin{equation}
\Delta^{ij,ab}_F(p)=-\frac{i\eta^{ij}\delta^{ab}}{p^2+i\epsilon}\,.
\end{equation}
The Feynman rule for the three-gluon vertex following from \eqref{tritr} is
\vspace{0.2cm}
\begin{equation}
\raisebox{-1.7cm}{
\includegraphics[width=0.29\textwidth]{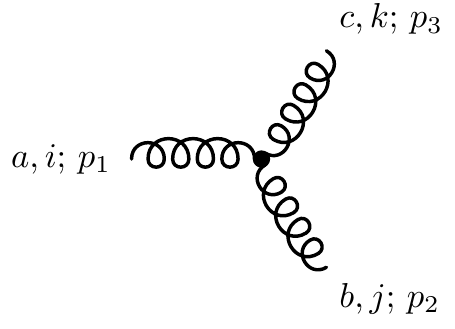}}\equiv ig_s\tilde f^{abc}V^{ijk}_{\textrm{YM}}(p_1,p_2,p_3)\,,
\end{equation}
\vspace{0.2cm}
\noindent with\footnote{Total antisymmetry under exchange of pairs of particles is guaranteed by momentum conservation but is not manifest in this form.}
\begin{equation}
V^{ijk}_{\textrm{YM}}(p_1,p_2,p_3)=\sqrt{2}\bigg[\eta^{ij}\bigg(\frac{p_{1+}}{p_{3+}}p^k_3-p^k_1\bigg)+\eta^{jk}\bigg(\frac{p_{2+}}{p_{1+}}p^i_1-p^i_2\bigg)+\eta^{ki}\bigg(\frac{p_{3+}}{p_{2+}}p^j_2-p^j_3\bigg)\bigg]\,,
\label{threeglue}
\end{equation}
where all momenta are outgoing, and the rescaled imaginary structure constants
\begin{equation}
\tilde f^{abc}\equiv i\sqrt{2}f^{abc}
\label{eq:ftilde}
\end{equation}
have been employed. 
The four-gluon interaction vertex derived from \eqref{quadritr}, again with all momenta outgoing, reads
\vspace{0.2cm}
\begin{eqnarray}
&&
\raisebox{-1.4cm}{\includegraphics[width=0.27\textwidth]{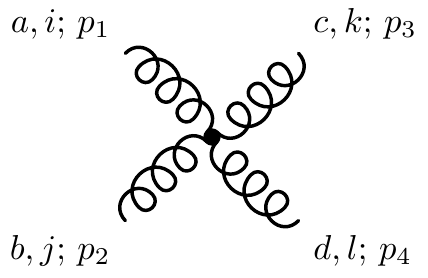}}
\equiv ig^2_s\bigg[\tilde f^{abe}\tilde f^{cde}V^{ijkl}_{s,\textrm{YM}}+\tilde f^{ace}\tilde f^{bde}V^{ijkl}_{t,\textrm{YM}}+\tilde f^{ade}\tilde f^{bce}V^{ijkl}_{u,\textrm{YM}}\bigg]
\end{eqnarray}
\vspace{0.2cm}
\noindent with\\[-0.8cm]
\begin{equation}
\begin{aligned}
V^{ijkl}_{s,\textrm{YM}}(p_1,p_2,p_3,p_4)&=\frac{1}{2}\bigg[\eta^{ij}\eta^{kl}\frac{(p_{1+}-p_{2+})(p_{4+}-p_{3+})}{(p_{1+}+p_{2+})^2}+\eta^{ik}\eta^{jl}-\eta^{il}\eta^{jk}\bigg]\,,\\
V^{ijkl}_{t,\textrm{YM}}(p_1,p_2,p_3,p_4)&=\frac{1}{2}\bigg[\eta^{ik}\eta^{jl}\frac{(p_{2+}-p_{4+})(p_{3+}-p_{1+})}{(p_{1+}+p_{3+})^2}+\eta^{ij}\eta^{kl}-\eta^{il}\eta^{jk}\bigg]\,,\\
V^{ijkl}_{u,\textrm{YM}}(p_1,p_2,p_3,p_4)&=\frac{1}{2}\bigg[\eta^{il}\eta^{jk}\frac{(p_{2+}-p_{3+})(p_{4+}-p_{1+})}{(p_{1+}+p_{4+})^2}+\eta^{ij}\eta^{kl}-\eta^{ik}\eta^{jl}\bigg]\,.
\end{aligned}
\end{equation}

\subsection{Einstein-Hilbert}
\label{ehlight}

The metric in the Einstein-Hilbert Lagrangian is expanded around the flat-space background
\begin{equation}
g_{\mu\nu}(x)=\eta_{\mu\nu}+\kappa h_{\mu\nu}(x)\,.
\end{equation}
The metric perturbation in light-cone gauge, $\ch_{\mu\nu}$, satisfies
\begin{equation}
n^{\mu}_+\ch_{\mu\nu}\equiv0\,,
\end{equation}
and can be decomposed as
\begin{equation}
\ch_{\mu\nu}=\frac{n_{+\mu}n_{+\nu}}{4}\,\ch_{--}+\frac{n_{+\mu}}{2}\ch_{-\nu_{\perp}}+\frac{n_{+\nu}}{2}\ch_{\mu_{\perp}-}+\ch_{\mu_{\perp}\nu_{\perp}}\,.
\end{equation}
Also here, only the transverse field components $\ch_{ij}$ are the propagating ones. We therefore integrate out the components $\ch_{-i}$, $\ch_{--}$, as well as the trace $\ch\equiv\ch^i_{\;i}$, by means of their equations of motion, which up to $\Lo(\kappa)$ have the solutions
\begin{align}
\ch &=\frac{\kappa}{2}\bigg[\ch_{ij}\ch^{ij}-\frac{1}{\p^2_+}(\p_+\ch_{ij}\p_+\ch^{ij})\bigg]\,,\\
\begin{split}
\ch_{-i} &=-\frac{2}{\p_+}\p^j\ch_{ij}+\kappa\bigg[
-\frac{\p_i}{\p^3_+}(\p_+\ch_{jk}\p_+\ch^{jk})
\\
&\quad\:\phantom{-\frac{2}{\p_+}\p^j\ch_{ij}+\kappa\bigg[}
+\frac{1}{\p^2_+}\bigg(-2\p^2_+\ch_{ij}\frac{\p_k}{\p_+}\ch^{jk}
+2\ch^{jk}\p_+\p_j\ch_{ik}
+\p_+\ch_{jk}\p_i\ch^{jk}\bigg)\bigg]\,.
\end{split}
\end{align}
Eliminating these components from the Lagrangian, we obtain up to trilinear order:
\begin{align}
\Ld^{\textrm{kin}}_{\textrm{EH}} &=\,\frac{1}{2}\p_{\mu}\ch_{ij}\p^{\mu}\ch^{ij}\,,\label{ehkin}\\
\begin{split}\label{eq:EHthreepointLCG}
\Ld^{(\kappa)}_{\textrm{EH}} &=\,\frac{\kappa}{2}\bigg[\ch_{ij}\p^2_+\ch^{ij}\frac{\p_k\p_l}{\p^2_+}\ch^{kl}-2\ch_{ij}\p_+\p^k\ch^{ij}\frac{\p^l}{\p_+}\ch_{kl}+\ch_{ij}\ch_{kl}\p^k\p^l\ch^{ij}\\
&\quad\:\phantom{\frac\kappa 2 \bigg[}
-2\ch_{il}\ch_{kj}\p^k\p^l\ch^{ij}
-4\ch_{il}\p_+\ch^{ij}\frac{\p^k\p^l}{\p_+}\ch_{kj}\bigg]\,.
\end{split}
\end{align}
We notice that by going back to a manifestly covariant form, 
the above expressions become
\begin{flalign}
\Ld^{\textrm{kin}}_{\textrm{EH}}=&\frac{1}{2}\p_{\alpha}\ch_{\mu\nu}\p^{\alpha}\ch^{\mu\nu}\,,\\
\Ld^{(\kappa)}_{\textrm{EH}}=&\frac{\kappa}{2}\Big[\ch_{\mu\nu}\ch_{\alpha\beta}\p^{\alpha}\p^{\beta}\ch^{\mu\nu}+2\ch_{\mu\nu}\p^{\alpha}\ch^{\mu\beta}\p^{\nu}\ch_{\alpha\beta}\Big],
\end{flalign}
which agrees with the kinetic and trilinear terms found in \cite{EHfromQCD}.

The graviton propagator in momentum space following from the graviton kinetic term \eqref{ehkin} reads:
\begin{equation}
\Delta^{ij,kl}_F(p)=\frac{i\eta^{ik}\eta^{jl}}{p^2+i\epsilon}\,.
\end{equation}
The three-graviton vertex Feynman rule derived from \eqref{eq:EHthreepointLCG} is:
\vspace{0.2cm}
\begin{equation}
\raisebox{-1.7cm}{\includegraphics[width=0.29\textwidth]{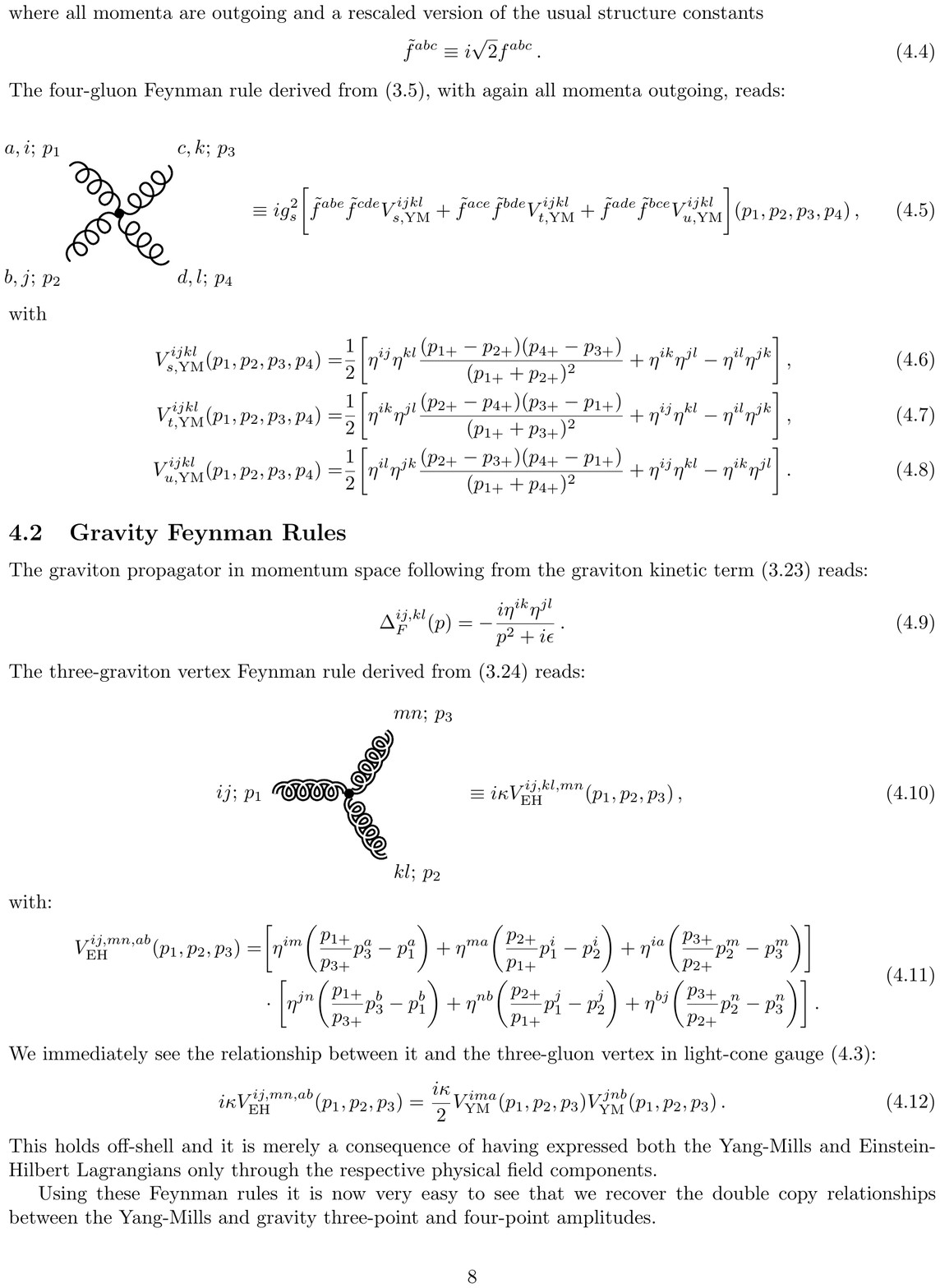}}\equiv -i\kappa V^{ij,kl,mn}_{\textrm{EH}}(p_1,p_2,p_3)\,,
\end{equation}
\vspace{-0.1cm}
\noindent with\footnote{Total symmetry under exchange of any pair of particles is guaranteed by the total antisymmetry of each square bracket.}
\begin{equation}\label{threegrav}
\begin{aligned}
V^{ij,kl,mn}_{\textrm{EH}}(p_1,p_2,p_3)&=\bigg[\eta^{ik}\bigg(\frac{p_{1+}}{p_{3+}}p^m_3-p^m_1\bigg)+\eta^{km}\bigg(\frac{p_{2+}}{p_{1+}}p^i_1-p^i_2\bigg)+\eta^{im}\bigg(\frac{p_{3+}}{p_{2+}}p^k_2-p^k_3\bigg)\bigg]\\
&\quad\times\bigg[\eta^{jl}\bigg(\frac{p_{1+}}{p_{3+}}p^n_3-p^n_1\bigg)+\eta^{ln}\bigg(\frac{p_{2+}}{p_{1+}}p^j_1-p^j_2\bigg)+\eta^{nj}\bigg(\frac{p_{3+}}{p_{2+}}p^l_2-p^l_3\bigg)\bigg]\,.
\end{aligned}
\end{equation}


\section{Mapping $\Ld_{\textrm{YM}}$ to $\Ld_{\textrm{EH}}$}
\label{mapping}

We are now ready to map the Yang-Mills Lagrangian to its colour-kinematics dual (CK-dual) Lagrangian. The double-copy field is  in light-cone gauge, $H_{+\mu}=0$, and therefore 
\begin{equation}
H_{\mu\nu}=\ch_{\mu\nu}+B_{\mu\nu}+C_{\mu\nu}\phi\,,
\label{decH}
\end{equation}
with $\ch_{+\mu}=B_{+\mu}=0$, and 
\begin{equation}
C_{\mu\nu}\equiv\frac{1}{2}\bigg(\eta_{\mu\nu}-\frac{n_{+\mu}\p_{\nu}+n_{+\nu}\p_{\mu}}{\p_+}\bigg)\,.
\end{equation}

\subsection{Kinetic term}
The kinetic term is mapped to
\begin{equation}
\Ld^{\textrm{kin}}_{\textrm{YM}}\mapsto\frac{1}{2}I\Big[\eta(\cA^{1\mu}\p_{\mu},\cA^{2\nu}\p_{\nu})\Big|\p_{\rho}\tilde\cA^1_i\p^{\rho}\tilde\cA^{2,i}\Big]=\frac{1}{2}I\Big[\cA^{1j}\cA^2_j\Big|\p_{\mu}\tilde\cA^1_i\p^{\mu}\tilde\cA^{2,i}\Big]\,,
\end{equation}
which, after taking the pairings, gives the CK-dual kinetic term
\begin{equation}
\Ld^{\textrm{kin}}_{\textrm{grav}}=\frac{1}{2}\p_{\mu}H_{ij}\p^{\mu}H^{ij}\,.
\end{equation}

\subsection{Trilinear graviton self-interaction}

The trace in \eqref{tritr} gets mapped to
\begin{equation}
\begin{split}
\tr(T^a[T^b,T^c]) &\mapsto-\frac{i}{4}\bigg[\cA^{1\mu}(\cA^{2\nu}\p_{\nu}\cA^3_{\mu}-\cA^{3\nu}\p_{\nu}\cA^2_{\mu})+\cA^{2\mu}(\cA^{3\nu}\p_{\nu}\cA^1_{\mu}-\cA^{1\nu}\p_{\nu}\cA^3_{\mu})\\
&\quad\:\phantom{-\frac i4\bigg[}+\cA^{3\mu}(\cA^{1\nu}\p_{\nu}\cA^2_{\mu}-\cA^{2\nu}\p_{\nu}\cA^1_{\mu})\bigg]\,,
\end{split}
\label{trilinmap}
\end{equation}
see \eqref{interf1}. Using the decomposition of $\cA_{\mu}$ in light-like directions \eqref{lightconedec}, and plugging in the linear part of \eqref{a-} for $\cA_-$, we find
\begin{equation}
\tr(T^a[T^b,T^c])\mapsto-\frac{i}{4}\sum_{\sigma\in S_3}\sign(\sigma)\bigg[\cA^{\sigma(1),i}\p_+\cA^{\sigma(2)}_i\frac{\p^j}{\p_+}\cA^{\sigma(3)}_j+\cA^{\sigma(1)}_i\cA^{\sigma(2)}_j\p^j\cA^{\sigma(3),i}\bigg]\,,
\end{equation}
where $\sigma$ denotes a permutation of the labels. The trilinear terms $\Ld^{(g_s)}_{\textrm{YM}}$ in \eqref{tritr} are therefore mapped to
\begin{eqnarray}
\frac{\kappa}{4}\sum_{\sigma\in S_3}\sign(\sigma)\,
I\bigg[\cA^{\sigma(1),i}\p_+\cA^{\sigma(2)}_i\frac{\p^j}{\p_+}\cA^{\sigma(3)}_j\!+\,\cA^{\sigma(1)}_i\cA^{\sigma(2)}_j\p^j\cA^{\sigma(3),i}\bigg|\,\tilde\cA^1_k\p_+\tilde\cA^{2,k}\frac{\p_l}{\p_+}\tilde\cA^{3,l}\!+
\,\tilde\cA^1_k\tilde\cA^2_l\p^l\tilde\cA^{3,k}\bigg].\;
\quad
\end{eqnarray}
Taking the pairings, simplifying the resulting expressions using integration by parts 
for $\partial_i$ and $\partial_+^{-1}$, and dropping total derivatives gives the double-copied trilinear Lagrangian
\begin{equation}
\begin{split}
\Ld^{(\kappa)}_{\textrm{grav}}&=\frac{\kappa}{2}\bigg[H_{ij}\p^2_+H^{ij}\frac{\p_k\p_l}{\p^2_+}H^{kl}-H^{ij}\p_+\p^{(k}H_{ij}\p^{l)}\frac{1}{\p_+}H_{kl}+H_{ij}H^{kl}\p_k\p_lH^{ij}\\
&\quad\:\phantom{\frac\kappa 2 \bigg[}
-2H_{il}H_{kj}\p^k\p^lH^{ij}-2H_{il}\p_+H^{ij}\frac{\p^k\p^l}{\p_+}H_{kj}-2H_{li}\p_+H^{ji}\frac{\p^k\p^l}{\p_+}H_{jk}\bigg]\,.
\end{split}
\label{dualtri}
\end{equation}
This expression can also be obtained by simply "squaring" the totally antisymmetrized colour-stripped trilinear Yang-Mills Lagrangian
\begin{equation}
\begin{split}
\Ld^{(\kappa)}_{\textrm{grav}}=\frac{\kappa}{4}\frac{1}{6}\sum_{\sigma,
\tilde\sigma\in S_3}&\sign(\sigma)\sign(\tilde\sigma)\,
I\bigg[\cA^{\sigma(1),i}\p_+\cA^{\sigma(2)}_i\frac{\p^j}{\p_+}\cA^{\sigma(3)}_j+\cA^{\sigma(1)}_i\cA^{\sigma(2)}_j\p^j\cA^{\sigma(3),i}\bigg|\\ 
&\qquad\phantom{\sign(\sigma)\sign(\sigma)} \tilde\cA^{\tilde\sigma(1)}_k\p_+\tilde\cA^{\tilde\sigma(2),k}\frac{\p_l}{\p_+}\tilde\cA^{\tilde\sigma(3),l}
+\tilde\cA^{\tilde\sigma(1)}_k\tilde\cA^{\tilde\sigma(2)}_l\p^l\tilde\cA^{\tilde\sigma(3),k}\bigg]\,,
\end{split}
\end{equation}
which is motivated by the known "squaring" relationship between the gluon and graviton three-point amplitudes. This is also the route taken in \cite{EHcubic} to derive the cubic Einstein-Hilbert Lagrangian via a double-copy construction. The authors of \cite{EHcubic}, however, needed to implement a non-linear field redefinition of $h_{\mu\nu}$ in the Einstein-Hilbert Lagrangian to show the equivalence of the graviton interaction terms obtained via double copy. Here it is immediately manifest, because we work only with physical degrees of freedom, and the result follows from the geometrically motivated mapping prescription.

We can again go back to a manifestly covariant form to find:
\begin{flalign}
\Ld^{\textrm{kin}}_{\textrm{grav}}=&\frac{1}{2}\p_{\alpha}H_{\mu\nu}\p^{\alpha}H^{\mu\nu}\,,\\
\Ld^{(\kappa)}_{\textrm{grav}}=&\frac{\kappa}{2}\bigg[H_{\mu\nu}H_{\alpha\beta}\p^{\alpha}\p^{\beta}H^{\mu\nu}+H^{\mu\nu}\p_{\mu}H^{\alpha\beta}\p_{\beta}H_{\alpha\nu}+H^{\mu\nu}\p_{\nu}H^{\alpha\beta}\p_{\alpha}H_{\mu\beta}\bigg]\,.
\end{flalign}
$\Ld^{(\kappa)}_{\textrm{grav}}$ is the same expression, up to the overall coupling, as constructed in \cite{cubicck} using the Noether procedure. However, unlike in \cite{cubicck}, we obtain this 
result by a rather straightforward mapping and pairing operation. 
We do not derive the CK-dual of the Yang-Mills four-point 
Lagrangian, since we cannot expect the linear CK mapping to 
work beyond the trilinear interaction.

\subsection{Eliminating the dilaton}
\label{elimdilaton}

We have now obtained the double copy of the Yang-Mills Lagrangian in light-cone gauge up to trilinear order, but we still have to check that this theory reproduces pure Einstein gravity at tree level. To this end, we substitute  the decomposition \eqref{decH} of $H_{\mu\nu}$ into the double-copied Lagrangian. The kinetic term becomes
\begin{equation}
\Ld^{\textrm{kin}}_{\textrm{grav}}=\frac{1}{2}\bigg[\p_{\mu}\ch_{ij}\p^{\mu}\ch^{ij}+\p_{\mu}B_{ij}\p^{\mu}B^{ij}+\frac{1}{2}\p_{\mu}\phi\p^{\mu}\phi\bigg]\,.
\label{dckin}
\end{equation}
Analogously, we obtain the full double-copied trilinear interaction Lagrangian by plugging \eqref{decH} into \eqref{dualtri}, see Appendix \ref{fullphystri}. Since the Kalb-Ramond field $B_{ij}$ is antisymmetric in its indices, it can only appear in pairs in the trilinear interaction terms, which is indeed what we find. Therefore these vertices cannot contribute to a tree-level scattering amplitude with only gravitons as external states. The trilinear terms that can contribute to tree-level pure graviton scattering amplitudes read
\begin{equation}
\begin{split}
\Ld^{(\kappa)}_{\ch^3,\ch^2\phi}&=\frac{\kappa}{2}\bigg[\ch_{ij}\p^2_+\ch^{ij}\frac{\p_k\p_l}{\p^2_+}\ch^{kl}-2\ch_{ij}\p_+\p^k\ch^{ij}\frac{\p^l}{\p_+}\ch_{kl}+\ch_{ij}\ch_{kl}\p^k\p^l\ch^{ij}-2\ch_{il}\ch_{kj}\p^k\p^l\ch^{ij}\\
&\qquad\:
-4\ch_{il}\p_+\ch^{ij}\frac{\p^k\p^l}{\p_+}\ch_{kj}+\frac{1}{2}\ch_{ij}\p^2_+\ch^{ij}\frac{\p_k\p^k}{\p^2_+}\phi-\ch_{ij}\p_+\p^k\ch^{ij}\frac{\p_k}{\p_+}\phi+\frac{1}{2}\phi\ch_{ij}\p_k\p^k\ch^{ij}\bigg]\,.
\end{split}
\label{finallag}
\end{equation}
We find the same trilinear graviton interactions as in Section \ref{ehlight}, but additional terms of the form $\ch\ch\phi$ appear, which suggest that the theory described by this Lagrangian deviates from pure Einstein gravity at tree level, starting from the four-point graviton scattering amplitude, which could have an internal dilaton line.  For instance, the following two $s$-channel $2\rightarrow2$ graviton 
scattering diagrams are possible in the CK-dual theory:

\begin{figure}[H]
\centering
\includegraphics[width=0.55\textwidth]{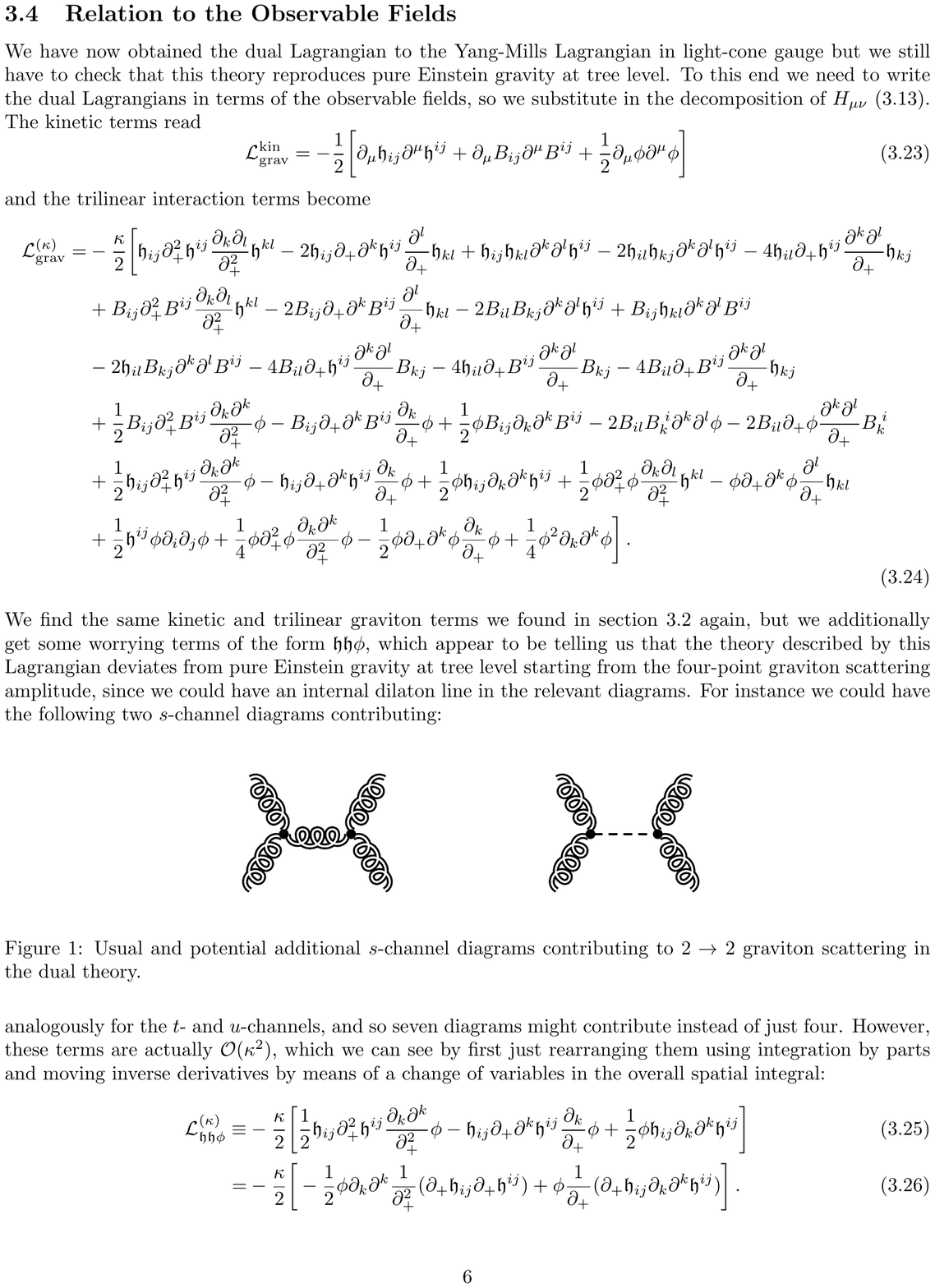}
\end{figure}

\noindent The dashed line represents the dilaton propagator. The analogous diagrams exist for the $t$- and $u$-channels. 
However, the dilaton exchange contribution can be shown to be equivalent 
to a four-point graviton contact vertex at $\Lo(\kappa^2)$ by applying the dilaton equation of motion, and hence is not relevant at the 
level of trilinear interactions considered here. To see this, 
we integrate by parts in the $\ch\ch\phi$ interaction Lagrangian 
to obtain
\begin{equation}
\Ld^{(\kappa)}_{\ch^2\phi}=\frac{\kappa}{2}\bigg[-\frac{1}{2}\phi\p_k\p^k\frac{1}{\p^2_+}(\p_+\ch_{ij}\p_+\ch^{ij})+\phi\frac{1}{\p_+}(\p_+\ch_{ij}\p_k\p^k\ch^{ij})\bigg]\,.
\label{trouble}
\end{equation} 
Using the linearized 
equation of motion of the graviton,
\begin{equation}
\p_k\p^k\ch_{ij}=-\p_+\p_-\ch_{ij}+\Lo(\kappa)\,,
\end{equation}
the last term in \eqref{trouble} is rewritten as 
\begin{align}
\phi\frac{1}{\p_+}(\p_+\ch_{ij}\p_k\p^k\ch^{ij})&=-\phi\frac{1}{\p_+}(\p_+\ch_{ij}\p_+\p_-\ch^{ij})+\Lo(\kappa)\nonumber\\
&=-\frac{1}{2}\phi\p_+\p_-\frac{1}{\p^2_+}(\p_+\ch_{ij}\p_+\ch^{ij})+\Lo(\kappa)\,.
\end{align}
Therefore, \eqref{trouble} becomes
\begin{align}
\Ld^{(\kappa)}_{\ch^2\phi}&=-\frac{\kappa}{4}\phi(\p_+\p_-+\p_k\p^k)\frac{1}{\p^2_+}(\p_+\ch_{ij}\p_+\ch^{ij})+\Lo(\kappa^2)\nonumber \\
&=-\frac{\kappa}{4}\frac{1}{\p^2_+}(\p_+\ch_{ij}\p_+\ch^{ij})\Box\phi+
\Lo(\kappa^2)\,.
\label{redef}\end{align}
The dilaton equation-of-motion $\Box\phi =0 + \Lo(\kappa)$ now 
allows us to push all offending terms to $\Lo(\kappa^2)$. 
Different from what was found in \cite{EHfromQCD}, this does \textit{not} generate non-local $\ch\ch\ch\ch$ couplings, where by "non-local" we mean containing $1/\Box$. We therefore have no $\ch\ch\phi$ interaction vertex at all, the remaining interaction terms between the graviton and the other fields are of the form $\ch\phi\phi$ and $\ch BB$ and do not contribute to tree-level scattering amplitudes with only gravitons in the initial and final states. \textit{The double-copied Lagrangian is therefore manifestly equivalent to the Einstein-Hilbert Lagrangian up to trilinear order when considering pure-graviton, tree-level scattering amplitudes.} 

In \cite{EHcubic} the dilaton was eliminated by coupling well-chosen unphysical sources to the BRST ghost fields, and it was observed that this can be done without constraining the source terms for the graviton. Here, since we work in light-cone gauge, which is ghost-free, we find that the dilaton decouples by itself from pure-graviton, tree-level scattering processes.

\subsection{Double-copied three-point vertex Feynman rule}
\label{gravfeyn}

We immediately see the squaring relationship between the three-gluon vertex \eqref{threeglue} and the three-graviton vertex \eqref{threegrav} in light-cone gauge:
\begin{equation}
\kappa V^{ij,kl,mn}_{\textrm{EH}}(p_1,p_2,p_3)=\frac{\kappa}{2}V^{ikm}_{\textrm{YM}}(p_1,p_2,p_3)V^{jln}_{\textrm{YM}}(p_1,p_2,p_3)\,.
\label{relation}
\end{equation}
This holds off-shell, and reflects the CK-duality between the trilinear gauge and gravitational Lagrangians in this gauge. It is merely a consequence of having expressed both the Yang-Mills and Einstein-Hilbert Lagrangians only through the respective physical field components. Since contracting the vertices with polarization vectors or tensors does not change anything, this relationship between the Feynman rules immediately implies the well-known "squaring" between gauge and gravity three-point amplitudes if the graviton polarization tensor is identified with the symmetric-traceless part of the direct product of two colour-stripped gluon polarization vectors in momentum space:
\begin{equation}
\ve_{ij}(k)\equiv\frac{1}{2}\Big[\ve_{(i}(k)\tilde\ve_{j)}(k)-\eta_{ij}\ve^l(k)\tilde\ve_l(k)\Big]\,.
\end{equation}


\section{Mapping matter Lagrangians}

The colour-kinematics duality mapping presented above not only applies to pure-gauge Lagrangians at trilinear order, but also to Lagrangians describing the coupling of matter to gauge fields. To illustrate this, we present as examples the application of it to the scalar and fermionic QCD Lagrangians to trilinear order. As stated in section \ref{rules} we will assume that the gauge field $\cA_{\mu}$ is in light-cone gauge, such that the resulting double-copy field after the mapping is as well.

\subsection{Scalar matter}

The Lagrangian of scalar QCD in light-cone gauge up to trilinear order reads
\begin{equation}
\Ld_{\textrm{sQCD}}=\p_{\mu}\phi^{\dagger}\p^{\mu}\phi-m^2\phi^{\dagger}\phi+ig_s\bigg[\phi^{\dagger}\cA_i\p^i\phi-\p^i\phi^{\dagger}\cA_i\phi+\p_+\phi^{\dagger}\frac{\p^i}{\p_+}\cA_i\phi-\phi^{\dagger}\frac{\p^i}{\p_+}\cA_i\p_+\phi\bigg]+\,\Lo(g^2_s)\,,
\label{eq:sQCD}
\end{equation}
where \eqref{a-} has been used to elminate $\cA_-$. The scalar field $\phi$ is a vector in colour space and its colour indices are contracted with the matrix indices $ij$ of the generators $T^a_{ij}$, which we left implicit above. Explicitly,
\begin{equation}
(\cA_{\mu}\phi)_i=\cA^a_{\mu}T^a_{ij}\phi_j\,.
\end{equation}
In covariant form, \eqref{eq:sQCD} reads
\begin{equation}
\Ld_{\textrm{sQCD}}=\p_{\mu}\phi^{\dagger}\p^{\mu}\phi-m^2\phi^{\dagger}\phi+ig_s\Big[\phi^{\dagger}\cA_{\mu}\p^{\mu}\phi-\p^{\mu}\phi^{\dagger}\cA_{\mu}\phi\Big]+\Lo(g^2_s)\,.
\label{sqcd}
\end{equation}

We now apply the mappings as described above. In the case of the scalar field, "colour-stripping" means that we remove the fundamental colour index $i$. Then we find 
\begin{flalign}
\Ld_{\textrm{sQCD}}\mapsto \;&\p_{\mu}\phi^*\p^{\mu}\phi-m^2\phi^*\phi-\kappa\bigg[\frac{\p^i\p^j}{\p^2_+}H_{ij}\p_+\phi^*\p_+\phi-\frac{\p^i}{\p_+}H_{ij}\p_+\phi^*\p^j\phi\nonumber\\
&-\frac{\p^j}{\p_+}H_{ij}\p^i\phi^*\p_+\phi+H_{ij}\p^i\phi^*\p^j\phi\bigg]+\Lo(\kappa^2)\,.
\label{Hscalar}
\end{flalign}
To obtain this form we eliminated $H_{-i}$, $H_{i-}$ and $H_{--}$, as the double-copy field in light-cone gauge satisfies the relations 
\begin{equation}
H_{-i}=-\frac{2}{\p_+}\p^jH_{ji}\,,\quad H_{i-}=-\frac{2}{\p_+}\p^jH_{ij}\,,\quad H_{--}=\frac{4}{\p^2_+}\p^i\p^jH_{ij}
\label{Hcomponents}
\end{equation}
at linear order, which it inherits from the gauge field in light-cone gauge and which ensure that $\p^{\mu}H_{\mu\nu}=0$. The mapped Lagrangian \eqref{Hscalar} describes a complex scalar field linearly coupled to the graviton, Kalb-Ramond and dilaton fields in light-cone gauge and can be brought back into the manifestly covariant form,
\begin{equation}
\Ld_{\textrm{sQCD}}\mapsto \p_{\mu}\phi^*\p^{\mu}\phi-m^2\phi^*\phi-\kappa H_{\mu\nu}\p^{\mu}\phi^*\p^{\nu}\phi\,.
\label{cov}
\end{equation}
Comparing this to \eqref{sqcd} we see that we could also have directly applied the mapping to the Lagrangian in covariant form and use that the double-copy field in light-cone gauge is transverse, $\p^{\mu}H_{\mu\nu}=0$, to bring it into the form \eqref{cov}. The pure Einstein gravity part of \eqref{Hscalar} is the same as
\begin{equation}
\Ld_{\phi}=\sqrt{-g}\Big[g^{\mu\nu}\p_{\mu}\phi^*\p_{\nu}\phi-m^2\phi^*\phi\Big]\,,
\label{covphi}
\end{equation}
with $g_{\mu\nu}=\eta_{\mu\nu}+\kappa\ch_{\mu\nu}$, expanded to linear order in $\kappa$ with $\ch_{\mu\nu}$ the graviton field in light-cone gauge and with unphysical components $\ch_{--}$, $\ch_{-i}$ and $\ch$ integrated out, such that only the two physical degrees of freedom propagate. Explicitly, 
\eqref{covphi} reads
\begin{equation}
\Ld_{\phi}=\p_{\mu}\phi^*\p^{\mu}\phi-m^2\phi^*\phi-\kappa\bigg[\frac{\p^i\p^j}{\p^2_+}\ch_{ij}\p_+\phi^*\p_+\phi-\frac{\p^i}{\p_+}\ch_{ij}[\p_+\phi^*\p^j\phi+\p_+\phi\p^j\phi^*]+\ch_{ij}\p^i\phi^*\p^j\phi\bigg]\,,
\end{equation}
or, brought back to manifestly covariant form,
\begin{equation}
 \Ld_{\phi}=\p_{\mu}\phi^*\p^{\mu}\phi-m^2\phi^*\phi-\kappa \ch_{\mu\nu}\p^{\mu}\phi^*\p^{\nu}\phi\,,
\end{equation}
where we used that the graviton field in light-cone gauge satisfies
\begin{equation}
\ch_{-i}=-\frac{2}{\p_+}\p^j\ch_{ij}+\Lo(\kappa)\,,\quad \ch_{--}=\frac{4}{\p^2_+}\p^i\p^j\ch_{ij}+\Lo(\kappa)\,,
\end{equation}
consistently with \eqref{Hcomponents}. Notice that in the fully covariant form of the expanded Lagrangian \eqref{covphi} without gauge-fixing also a term proportional to the trace $h$ at $\Lo(\kappa)$ appears and $\p^{\mu}h_{\mu\nu}$ does not vanish even at the linear level, thus obscuring the link to \eqref{sqcd}. As for the case of pure Yang-Mills theory studied in Section \ref{mapping} we see that the double-copy relation at the Lagrangian level becomes manifest once only the physical degrees of freedom are left in the description.

\subsection{Dirac fermion matter}
Applied to the fermionic QCD Lagrangian
\begin{flalign}
\Ld_{\textrm{QCD}}=&\,\bar\psi(i \gamma^{\mu}\p_{\mu}-m)\psi-g_s\bar\psi\bigg[\frac{\p^i}{\p_+}\cA_i\slashed{n}_+-\gamma^i\cA_i\bigg]\psi\\
=& \,\bar\psi(i\gamma^{\mu}\p_{\mu}-m)\psi+g_s\bar\psi\gamma^{\mu}\cA_{\mu}\psi\,,\label{fqcd}
\end{flalign}
the mapping results in 
\begin{eqnarray}
\Ld_{\textrm{QCD}}& \mapsto &
\bar\psi(i\gamma^{\mu}\p_{\mu}-m)\psi\nonumber\\
&& -\,\frac{i\kappa}{2}\bar\psi\bigg[\frac{\p^i\p^j}{\p^2_+}H_{ij}\slashed{n}_+\p_+-\frac{\p^i}{\p_+}H_{ij}\slashed{n}_+\p^j-\frac{\p^i}{\p_+}H_{ji}\gamma^j\p_++H_{ij}\gamma^i\p^j\bigg]\psi\\
&=&\bar\psi(i\gamma^{\mu}\p_{\mu}-m)\psi-\frac{i\kappa}{2}H_{\mu\nu}\bar\psi\gamma^{\mu}\p^{\nu}\psi\,.\label{Hfermion}
\end{eqnarray}
Also here we implicitly colour-stripped the quark field $\psi_i$ of its fundamental index $i$ during the mapping. The pure Einstein gravity part of this Lagrangian is the same as the Lagrangian of a Dirac fermion on a curved background metric $g_{\mu\nu}=\eta_{\mu\nu}+\kappa\ch_{\mu\nu}$, expanded to linear order in $\kappa$, with $\ch_{\mu\nu}$ the graviton field in light-cone gauge and with unphysical components integrated out, as above, which eliminates terms proportional to $\ch$ and $\p^{\mu}\ch_{\mu\nu}$ and leaves only, in manifestly covariant form,
\begin{equation}
\Ld_{\psi}=\bar\psi(i\gamma^{\mu}\p_{\mu}-m)\psi-\frac{i\kappa}{2}\ch_{\mu\nu}\bar\psi\gamma^{\mu}\p^{\nu}\psi+\Lo(\kappa^2)\,.\label{covf}
\end{equation} 
We could have again applied the mapping directly to the manifestly covariant Lagrangian \eqref{fqcd} to obtain \eqref{Hfermion}. The double-copy link between the gauge and gravitational Lagrangians \eqref{fqcd}, \eqref{covf} becomes obvious once only the physical degrees of freedom are left in the description, but is obscured in their most general form.


\section{Next-to-soft graviton theorem from colour-kinematics dual SCET} 

We now apply our colour-kinematics duality mapping to the soft-collinear effective theory (SCET) of QCD to derive the fermionic soft and next-to-soft graviton theorem from the corresponding gauge theory results.\footnote{For a derivation of the gravitational soft theorem from the soft-collinear effective theory of gravity, see \cite{Beneke:2021umj}.}

\subsection{Gravitational soft theorem}

The soft theorem in gauge theory is known as the Low-Burnett-Kroll (LBK) theorem \cite{L,BK}, which states that, given a non-radiative scattering amplitude $\A_\textrm{full}$ of $n$ energetic particles, the 
amplitude $\A_A$ for the emission of an additional soft gauge boson 
has the universal form
\begin{equation}
\A_A=-g_s\sum_{i=1}^n \psi_i\,T^a_i\bigg[\frac{p_i\cdot\ve^a(k)}{p_i\cdot k}+\frac{k_{\nu}\ve^a_{\mu}(k)J^{\mu\nu}_i}{p_i\cdot k}\bigg]\A_i\,,
\label{gaugesoft}
\end{equation}
where  $\psi_i(p_i)$ refers to the polarization of the energetic particle 
with momentum $p_i$. The gravitational next-to-soft theorem has the universal form \cite{weinberg,CS}  
 \begin{equation}
 \A_h=\frac{\kappa}{2}\sum_{i=1}^n\psi_i\bigg[\frac{\ve_{\mu\nu}(k)p^{\mu}_ip^{\nu}_i}{p_i\cdot k}+\frac{\ve_{\mu\nu}(k)p^{\mu}_ik_{\rho}J^{\nu\rho}_i}{p_i\cdot k}+\Lo(k)\bigg]\A_i\,,
 \label{gravsoft}
 \end{equation}
with $\A_h$ the gravitation-radiation amplitude. In both cases, $\A_i$ is the non-radiative amplitude with the $i$-th 
polarization function stripped off,
\begin{equation}
\A_{\textrm{full}}=\psi_i\A_i \qquad\mbox{ (no sum over $i$)}\,,
\label{eq:afull}
\end{equation} 
and all particles are taken to be outgoing. This makes explicit that the soft factor in the square brackets in \eqref{gaugesoft}, \eqref{gravsoft} acts only on the amplitude without the polarization functions, which will be important later on. Writing the soft theorems in this manner emphasizes that they really stem from the presence of soft momenta in the scattering process, not from the properties of the external legs. Further, 
\begin{equation}
J^{\mu\nu}_i\equiv L^{\mu\nu}_i+S^{\mu\nu}_i
\end{equation}
is the angular momentum operator, separated into the orbital angular momentum operator
\begin{equation}
L^{\mu\nu}_i\equiv p^{[\mu}_i\frac{\p}{\p p_{\nu]i}}\,,
\end{equation}
and the spin operator $S^{\mu\nu}_i$, which takes into account the 
intrinsic angular momentum of the particles, in the appropriate representation of the Lorentz group. 
Since we are interested in the fermionic version of the soft 
theorem, the spin operator reads 
\begin{equation}
S^{\mu\nu}_iu(p_i)=\frac{1}{4}[\gamma^{\mu},\gamma^{\nu}]u(p_i)\,,\qquad \bar u(p_i)S^{\mu\nu}_i=-\frac{1}{4}\bar u(p_i)[\gamma^{\mu},\gamma^{\nu}]
\end{equation}
when acting on Dirac spinors $u(p)$ and their adjoint $\bar u(p)\equiv u^{\dagger}(p)\gamma^0$ in $\A_\textrm{full}$. We can therefore summarize the action of $S^{\mu\nu}_i$ on the fermionic amplitude as
\begin{equation}
S^{\mu\nu}_i\rightarrow\frac{\eta_i}{4}[\gamma^{\mu},\gamma^{\nu}]\,,
\end{equation}
where, following the notation of \cite{weinberg},
\begin{equation}
\eta_i\equiv\begin{cases}
+1\,,&i\textrm{-th particle outgoing}\\
-1\,,&i\textrm{-th particle incoming}
\end{cases}\,.
\end{equation}

Comparing the soft factors in the gauge and gravitational cases, we already see the colour-kinematics duality prescription manifest itself, since \eqref{gravsoft} can be obtained from \eqref{gaugesoft} by the replacements 
\begin{equation}
\ve^a_{\mu}(k)T^a\rightarrow-\ve_{\mu\nu}(k)p^{\nu}_i\,,\quad g_s\rightarrow\frac{\kappa}{2}\,.
\end{equation}
We will show that this follows from the corresponding  relation for the soft emission effective Lagrangians.

\subsection{Effective Lagrangian from colour-kinematics duality}

We now apply the mappings defined earlier to the SCET QCD Lagrangian in the position-space formalism given in \cite{Beneke:2002ni,Beneke:2002ph},\footnote{See Appendix \ref{DCSCET} for a brief review of the basics of SCET QCD and the explicit expressions for the Lagrangian.} and obtain an effective description for a fermion coupled to gravity.
To construct the Lagrangian, we apply the mapping \eqref{finalform} to the soft gluon field,
\begin{equation}
A^a_{s\mu}T^a\mapsto-iS_{\mu\nu}\p^{\nu}\,,
\end{equation}
where
\begin{equation}
S_{\mu\nu}=s_{\mu\nu}+B_{s,\mu\nu}+C_{\mu\nu}\phi_s\,,
\end{equation}
with the soft graviton, Kalb-Ramond, and dilaton fields. We assume $S_{\mu\nu}$ to be in a transverse gauge as well, such that $\p^{\mu}S_{\mu\nu}=0$. 
The $\lambda$-scaling of the soft double-copy field is 
$S_{\mu\nu}\sim\lambda^2$ \cite{SCETgravity}. 

Using the rules established in Section~\ref{rules}, the soft field-strength tensor is mapped to
\begin{equation}
F^a_{s,\mu\nu}T^a\mapsto-iK_{s,\mu\nu\rho}\p^{\rho}\,,
\end{equation}
with
\begin{equation}
K_{s, \mu\nu\rho}\equiv\p_{\mu}S_{\nu\rho}-\p_{\nu}S_{\mu\rho}+\frac{\kappa}{4}\bigg[S_{\mu\sigma}\p^{\sigma}S_{\nu\rho}-S_{\nu\sigma}\p^{\sigma}S_{\mu\rho}\bigg]\,,
\end{equation} 
which satisfies $\p^{\rho}K_{\mu\nu\rho}=0$ to linear order, 
as sufficient. The gauge-covariant derivative for the fermion field, $D_{\mu}$, is mapped to
\begin{equation}
iD_{\mu}=i\p_{\mu}+g_sA^a_{\mu}T^a\mapsto iD_{\mu}\equiv i\p_{\mu}-\frac{i\kappa}{2}S_{\mu\nu}(x)\p^{\nu}\,.
\end{equation}

Applying these mappings to the soft-collinear interaction Lagrangians $\Ld^{(m)}_{\xi}$, $m=0,1,2$ up to order $\mathcal{O}(\lambda^2)$ (explicit expressions given in Appendix \ref{DCSCET}) and fixing collinear light-cone gauge, yields the effective Lagrangian for the emission of a soft graviton from an energetic Dirac fermion in the form
\begin{align}
\Ld^{(0)}_D&=-\frac{\kappa}{4}\bar\xi n^{\mu}_-n^{\nu}_-S_{\mu\nu}(n_+\cdot i\p)\frac{\slashed{n}_+}{2}\xi\,,\label{l0dc}\\
\Ld^{(1)}_D&=-\frac{\kappa}{4}\bar\xi\Big(x^{\mu}_{\perp}n^{\nu}_-n^{\rho}_-K_{s,\mu\nu\rho}(n_+\cdot i\p)+2n^{\mu}_-S_{\mu\nu}i\p^{\nu}_{\perp}\Big)\frac{\slashed{n}_+}{2}\xi\,,\label{l1dc}\\
\begin{split}\label{l2dc}
\Ld^{(2)}_D&=\bar\xi\bigg[-\frac{\kappa}{8}\bigg((n_-\cdot x)n^{\mu}_+n^{\nu}_-n^{\rho}_-K_{s,\mu\nu\rho}(n_+\cdot i\p)+x^{\mu}_{\perp}x_{\perp\rho}n^{\nu}_-n^{\sigma}_-\Big([\p^{\rho},K_{s,\mu\nu\sigma}(n_+\cdot i\p)]\\
&\qquad\quad\phantom{-\frac\kappa8\bigg[} +\frac{\kappa}{4}\Big[S^{\rho\alpha}i\p_{\alpha} K_{s,\mu\nu\sigma}(n_+\cdot i\p)-K_{s,\mu\nu\sigma}(n_+\cdot i\p)S^{\rho\alpha}i\p_{\alpha}\Big]\Big)\bigg)\\
&\quad\:-\frac{\kappa}{8}\bigg(i\slashed \p_{\perp}\frac{1}{in_+\cdot \p}x^{\mu}_{\perp}\gamma^{\nu}_{\perp}n^{\rho}_-K_{s,\mu\nu\rho}(n_+\cdot i\p)+x^{\mu}_{\perp}\gamma^{\nu}_{\perp}n^{\rho}_-K_{s,\mu\nu\rho}(n_+\cdot i\p)\frac{1}{in_+\cdot \p}i\slashed \p_{\perp}\bigg)\\
&\quad\:-\frac{i\kappa}{4}\Big(n^{\mu}_-n^{\nu}_+S_{\mu\nu}(n_-\cdot i\p)+2n^{\nu}_-x^{\mu}_{\perp}K_{s,\mu\nu\rho}i\p^{\rho}_{\perp}\Big)\bigg]\frac{\slashed{n}_+}{2}\xi\,.
\end{split}
\end{align}
The Feynman rules for the interaction vertices between collinear fermions and a soft graviton up to $\Lo(\lambda^2)$ are given in Appendix \ref{sec:gravitySCETFeynmanrule}. For completeness, we also show the non-linear terms in $S_{\mu\nu}$, which follow from the prescription. However, as discussed below, since we expect that the linear mapping prescription should receive higher-order corrections, only the terms linear in $S_{\mu\nu}$ in the above Lagrangian should be taken for granted. These suffice to derive the soft theorem, which refers to single-graviton emission. 

\subsection{Deriving the soft theorem}
\label{comp}

We now derive the soft theorem from the soft-collinear gravity Lagrangian including subleading powers, as obtained from the CK-duality mapping. For $N$ energetic particles with momenta $p_i$, well-separated in angle, the Lagrangian takes the form
\begin{equation}
    \mathcal{L}_\mathrm{SCET} = \sum_{i=1}^N \mathcal{L}_{D,i}(\psi_i,\psi_s) + \mathcal{L}_s(\psi_s)\,,
\end{equation}
with $\mathcal{L}_{D,i}$ the CK-dual soft-collinear Lagrangian for the $i$-th collinear sector, given by \eqref{l0dc} to \eqref{l2dc} with the replacement $n_+\rightarrow n_{i+}$, and $\mathcal{L}_s$ the purely soft Lagrangian. Notably, $i$-collinear fields only interact among themselves and 
with soft fields, but not with collinear fields of another direction. To simplify the computation of the radiative amplitude, we adopt a 
frame in which the external momenta $p_i^\mu$ 
are aligned with the light-like reference 
vectors $n^{\mu}_{i-}$, $i=1,...,n$, which implies $p_{i\perp}=0$, 
$p_{i-}=0$. This choice can always be made. 

In addition to the Lagrangian, which describes soft 
emission, we need source operators, which generate the 
energetic particles, and possibly soft particles. A generic operator in QCD SCET takes the form \cite{Beneke:2017ztn,Beneke:2018rbh}
\begin{gather}
    \mathcal{J} = \int\der t\: C(\{t_{i_k}\}) J_s(0) \prod_{i=1}^N J_i(t_{i_1},t_{i_2},\dots)\,,
\end{gather}
where $\der t =\prod_{ik}\der t_{i_k}$, $C$ is the matching coefficient, $J_s$ is a purely soft building block, and $J_i$ are the collinear building blocks. An important observation is that up to $\mathcal{O}(\lambda^2)$, $J_s(0)=1$ by gauge invariance. Hence, to derive the (next-to-)~soft theorem, only collinear source building blocks are available, and the soft gluon must arise from a soft emission Lagrangian vertex.  For more details on the operator basis, we refer to \cite{Beneke:2017ztn,Beneke:2018rbh}.

Applying the mappings of Section~\ref{rules} to this operator basis is trivial in collinear light-cone gauge, which sets all collinear Wilson lines $W_c=1$. The relevant collinear building block is the outgoing spinor field $\chi^{\dagger}_i = \xi^{\dagger}_i$ and its transverse derivatives 
up to $\Lo(\lambda^2)$,
\begin{equation}
\begin{aligned}
&\Lo(\lambda^0): & &J^{A0}_{\chi_{i}}(t_i) = \chi^{\dagger}_{i}(t_in_{i+})\,,\\
&\Lo(\lambda^1): & &J^{A1\mu}_{\p\chi_{i}}(t_i) = i\p^{\mu}_{i\perp}\chi^{\dagger}_{i}(t_in_{i+})\,,\\
&\Lo(\lambda^2): & &J^{A2\mu\nu}_{\p^2\chi_{i}}(t_i) = i\p^{\mu}_{i\perp}i\p^{\nu}_{i\perp}\chi^{\dagger}_{i}(t_in_{i+})\,,
\end{aligned}
\end{equation}
in complete analogy to the QCD case. 
Current operators involving more than one collinear building block in a single collinear sector, such as the $Bn$- or $Cn$-type operators \cite{Beneke:2017ztn,Beneke:2018rbh}, cannot contribute here, since we assume that we have exactly one collinear particle in each sector, and work at tree level.
Notably, up to $\mathcal{O}(\lambda^2)$, there are also no soft building blocks in the gravitational case.

In addition to the subleading contributions due to $A1, A2$ currents, power-suppressed next-to-soft terms arise from time-ordered products of the current operators with the subleading Lagrangian, 
\begin{equation}
    J^{Tn}(t_i) = i\int\der^4x\: T\{ J^{A0}(t_i)\mathcal{L}^{(n)}_{\xi,i}(x)\}\,,
\end{equation}
representing subleading interaction vertices. As we will see below, the entire soft theorem stems from such time-ordered products.

With the notation set up, we calculate the soft graviton emission amplitude in SCET gravity. Given the adopted frame, the non-radiative amplitude is simply
        \begin{equation} 
        	\A_{\textrm{full}} = \bra{q(p_1)\dots q(p_n)}\int \der t\; C^{A0}(\{t_i\}) \prod_{j=1}^nJ^{A0}_{\chi_j}(t_j)\ket{0}\bigg|_{\substack{\textrm{tree}\,,\\ p^{\mu}_{i\perp}=0}}\,.
        \end{equation}  
We take all particles to be outgoing. This evaluates to
\begin{equation}
\A_{\textrm{full}}=\bar\xi_{1...n}C^{A0}(\{p_{i+}\})\,,
\label{match}
\end{equation}
where
\begin{equation}
\bar\xi_{1...n}\equiv\bar\xi_1(p_1)...\bar\xi_n(p_n)\,,
\label{xi}
\end{equation}
and we suppress both the spinor indices of the $\xi_i$ and of the Fourier-transformed Wilson coefficient, which is defined by
        \begin{equation} 
        	C(t_i) = \int \der p_i \,e^{-i t_i p_i}C(p_i)\,.
        \end{equation} 
We use the short-hand notations $p_{i\pm}\equiv n_{i\pm}\cdot p$ and $\ve_{i\pm,j\pm}\equiv n^{\mu}_{i\pm}n^{\nu}_{j\pm}\ve_{\mu\nu}$, and, for the sake of conciseness, abbreviate $\bra{q(p_1)\dots q(p_n)}$ to $\bra{q}$.
        
The $C^{A1}$ coefficient is completely determined in terms of 
$C^{A0}$ by reparametrization invariance constraints. For all-outgoing fermions, it is given by
        \begin{equation}
    \begin{split}
    C^{A1\mu}_i(\{p_{k+}\})=&\bigg[-\frac{\gamma^{\mu}_{i\perp}}{p_{i+}}\frac{\slashed{n}_{i+}}{2}-\sum_{j\neq \,i}\frac{2n^{\mu}_{j-}}{n_{i-}\cdot n_{j-}}\frac{\p}{\p p_{i+}}\bigg]C^{A0}(\{p_{k+}\})\\
    \equiv& \,C^{A1\mu}_{i,\textrm{spinor}}(\{p_{k+}\})+C^{A1\mu}_{i,\textrm{coll.}}(\{p_{k+}\})\,.
    \end{split}
    \label{A1}
    \end{equation}
The first summand contributes to the reparametrization invariant completion of the two-component spinor $\xi$ \cite{rpispinor} and the second summand holds for any type of collinear particle, and represents the generalization  \cite{BGS} to $n$ distinct collinear directions of the relation given in \cite{A1}. In the full theory this term stems from the expansion of the (stripped) hard amplitude in the SCET power-counting, as will be seen below.
    
With our choice of momenta, where $p_{i\perp} = 0$, there are only three non-vanishing contributions to the emission of a soft graviton,
\begin{equation}
\begin{split}
\A_h=\,&\bra{q s_{\mu\nu}(k)}\int\der t\;C^{A0}(\{t_i\})\prod_{j=1}^nJ^{A0}_{\chi_{j}}(t_j)\ket{0}\\
&+\bra{q s_{\mu\nu}(k)}i\sum_{k=1}^n\int\der t\der^4z\;C^{A1}_{k\mu}(\{t_i\})\,T\{J^{A1\mu}_{\p\chi_{k}}(t_k)\Ld^{(1)}_{D,k}(z)\}\prod_{j\neq k}J^{A0}_{\chi_{j}}(t_j)\ket{0}\\
&+\bra{q s_{\mu\nu}(k)}i\sum_{k=1}^n\int\der t\der^4z\;C^{A0}(\{t_i\})\,T\{J^{A0}_{\chi_{k}}(t_k)\Ld^{(2)}_{D,k}(z)\}\prod_{j\neq k}J^{A0}_{\chi_{j}}(t_j)\ket{0}\,,
\end{split}
\label{radamp}
\end{equation}
where the first line of \eqref{radamp} contributes at $\Lo(\lambda^0)$ and the last two lines contribute at $\Lo(\lambda^2)$. Due to our frame choice, there is no contribution to the soft theorem at $\Lo(\lambda)$, which is consistent with the expansion in the soft momentum.
The second line of \eqref{radamp} is \textit{not} vanishing: 
the momentum space expression of $X^{\mu}_{k\perp}$ contained in 
$\Ld^{(1)}_{D,k}$, see Appendix \ref{sec:gravitySCETFeynmanrule},
acts on the contraction of the current operator $J^{A1\mu}_{\p\chi_k}$ with the $\xi$ field from 
$\Ld^{(1)}_{D,k}$ \textit{before} setting $p^{\mu}_{i\perp}=0$. 

To evaluate the summands in \eqref{radamp}, we can consider one collinear direction at a time and add up all the contributions at the end. For the $i$-th fermion, using the Feynman rules in Appendix \ref{sec:gravitySCETFeynmanrule}, we find at $\Lo(\lambda^0)$
\begin{equation}
\bra{q s_{\mu\nu}(k)}\int\der t\;C^{A0}(\{t_i\})\prod_{j=1}^nJ^{A0}_{\chi_{j}}(t_j)\ket{0}=\frac{\kappa}{2}\frac{\ve_{i-,i-}p_{i+}}{2k_{i-}}\bar\xi_{1...n}\,C^{A0}(\{p_{j+}\})\,.
\label{A0}
\end{equation}
At $\Lo(\lambda^2)$ we obtain the two pieces
\begin{align}
&\bra{q s_{\mu\nu}(k)}i\sum_{k=1}^n\int\der t\der^4z\;C^{A1\mu}(\{t_i\})\,T\{J^{A1}_{\mu\p\chi_{k}}(t_k)\Ld^{(1)}_{D,k}(z)\}\prod_{j\neq k}J^{A0}_{\chi_{j}}(t_j)\ket{0}\nonumber\\
&=\frac{\kappa}{2}\bar\xi_{1...n}\bigg[\frac{\ve_{i-,\nu}k_{\rho}\gamma^{[\nu}_{i\perp}n^{\rho]}_{i-}}{2k_{i-}}\frac{\slashed{n}_{i+}}{2}
-\frac{p_{i+}(\ve_{i-,i-}k_{\mu i\perp}-\ve_{i-,\mu_{i\perp}}k_{i-})}{2k_{i-}}\sum_{k\neq i}\frac{2n^{\mu}_{k-}}{n_{i-}\cdot n_{k-}}\frac{\p}{\p p_{i+}}\bigg]C^{A0}(\{p_{j+}\})\,,
\label{A21}\end{align}
and 
\begin{equation}
\begin{split}
&\bra{q s_{\mu\nu}(k)}i\sum_{k=1}^n\int\der t\der^4z\;C^{A0}(\{t_i\})\,T\{J^{A0}_{\chi_{k}}(t_k)\Ld^{(2)}_{D,k}(z)\}\prod_{j\neq k}J^{A0}_{\chi_{j}}(t_j)\ket{0}\\
&=\frac{\kappa}{2}\bar\xi_{1...n}\bigg[\frac{k_{i+}\ve_{i-,i-}-k_{i-}\ve_{i+,i-}}{2k_{i-}}\bigg(\frac{1}{2}+p_{i+}\frac{\p}{\p p_{i+}}\bigg)+\frac{\ve_{i-,\mu}[\gamma^{\mu}_{i\perp},\slashed{k}_{i\perp}]}{4k_{i-}}\bigg]C^{A0}(\{p_{j+}\})\,.
\end{split}
\label{A22}
\end{equation}
To put \eqref{A22} into the given form, we have used the fact that the on-shell graviton polarization tensor $\ve_{\mu\nu}(k)$ is transverse and traceless,
\begin{equation}
k^{\mu}\ve_{\mu\nu}(k)=0\,,\quad\ve^{\mu}_{\;\mu}(k)=0\,.
\end{equation}
In conclusion, we find the following amplitude for the soft emission of a graviton off of the $i$-th leg:
\begin{equation}
\begin{split}\label{softfromscet}
    \mathcal{A}_{h,\textrm{ leg }i} &= \frac{\kappa}{2}\bar\xi_{1...n}\bigg[
    \frac{\varepsilon_{i-,i-}p_{i+}}{2 k_{i-}}C^{A0}(\{p_j\})
    + \frac{\varepsilon_{i-,i-}k_{i+} - \varepsilon_{i-,i+}k_{i-}}{2k_{i-}}p_{i+}\frac{\partial}{\partial p_{i+}}C^{A0}(\{p_j\})\\
    &\quad\:\phantom{-\frac\kappa2\bar\xi_{1\dots n}\bigg[}
    - \frac{\varepsilon_{i-,i-}k_{\mu_{i\perp}} - \varepsilon_{i-,\mu_{i\perp}}k_{i-}}{2k_{i-}}p_{i+}\sum_{k\neq i}\frac{2 n_{k-}^\mu}{n_{i-}\cdot n_{k-}} \frac{\partial}{\partial p_{i+}}C^{A0}(\{p_j\})\\
    &\quad\:\phantom{-\frac\kappa2\bar\xi_{1\dots n}\bigg[}
    + \frac{\varepsilon_{i-,[i-} k_{i+]} + \varepsilon_{i-,\nu}k_\rho \gamma_{i\perp}^{[\nu} n_{i-}^{\rho]}\slashed{n}_{i+} + \varepsilon_{i-,\nu}[\gamma_{i\perp}^\nu,\slashed{k}_{i\perp}]}{4 k_{i-}} C^{A0}(\{p_j\})\bigg]\,.
    \end{split}
\end{equation}
We can now compare this to the soft theorem. Inserting the 
SCET decomposition of vectors in $n_\pm^\mu$ and transverse 
components into \eqref{gravsoft} with only outgoing particles, 
we find
\begin{align}
\A_h&=\frac{\kappa}{2}\sum_{i=1}^n\bar\xi_i\bigg[\frac{\ve_{i-,i-}p_{i+}}{2k_{i-}}+\frac{\ve_{i-,i-}k_{i+}-\ve_{i-,i+}k_{i-}}{2k_{i-}}p_{i+}\frac{\p}{\p p_{i+}}+\frac{\ve_{i-,i-}k^{\nu}_{i\perp}-\ve^{\nu}_{\;i-}k_{i-}}{2k_{i-}}p_{i+}\frac{\p}{\p p^{\nu}_{i\perp}}\nonumber\\
&\quad\:\phantom{-\frac\kappa2\sum\bar\xi\bigg[}
+\frac{\ve_{i-,[i-}k_{i+]}+\ve_{i-,\nu}k_{\rho}\gamma^{[\nu}_{i\perp}n^{\rho]}_{i-}\slashed{n}_{i+}+\ve_{i-,\nu}[\gamma^{\nu}_{i\perp},\slashed{k}_{i\perp}]}{4k_{i-}}\bigg]\A_i+\Lo(\lambda^3)\,.\label{softfer}
\end{align}
Even though we set $p^{\mu}_{\perp i}=0$, $p_{i-}=0$ for the external momenta, we must do so \textit{after} taking the momentum derivatives, and, therefore, the last term in the first line of \eqref{softfer} is non-vanishing. It naively looks like an $\Lo(\lambda)$-term, however, it actually counts as $\Lo(\lambda^2)$, since the $p_{i\perp}$-derivatives always cancel factors of $p^{\mu}_{i\perp}$ in the non-radiative amplitude.

We have already matched the $A0$-coefficient in \eqref{match}, so we immediately see that the first and last lines of \eqref{softfromscet} reproduce every term but the last one in the first line of \eqref{softfer}. This last term must be reproduced by the second line of \eqref{softfromscet}. It therefore remains to show that
\begin{equation}
\bar\xi_i\,\frac{\p}{\p p_{\mu i\perp}}\A_i\bigg|_{p^{\mu}_{i\perp}=0}
= - \bar\xi_{1...n}\,\sum_{k\neq i}\frac{2 n_{k-}^\mu}{n_{i-}\cdot n_{k-}} \frac{\partial}{\partial p_{i+}}C^{A0}(\{p_{j+}\})\,.
\label{eq:perprel}
\end{equation}
To see this, we observe that the non-radiative amplitude in momentum space, \textit{before} setting $p^{\mu}_{i\perp}=0$, is given by
\begin{equation}
\A_{\textrm{full}}=\bar\xi_{1...n}\bigg[C^{A0}(\{p_{i+}\})+\sum_{i=1}^np^{\mu}_{i\perp}C^{A1}_{\mu i}(\{p_{j+}\})+\Lo(\lambda^2)\bigg]
\end{equation}
with the same $A0$- and $A1$-coefficients as given above. 
From \eqref{eq:afull}, we therefore find 
\begin{equation}
\left(\frac{\p}{\p p_{\mu i\perp}}\bar\psi_i(p_i)\right)\A_i
+\bar\psi_i(p_i)\frac{\p}{\p p_{\mu i\perp}}\A_i 
=\bar\xi_{1...n}\,C^{A1\mu}_ i(\{p_{j+}\})
+\Lo(\lambda^2)\,.
\label{eq:relperp}
\end{equation}
The full Dirac spinor is related to the SCET collinear 
spinor by \cite{rpispinor} 
\begin{equation}
\psi_i(p_i)=\bigg[1+\frac{\slashed{p}_{i\perp}\slashed{n}_{i+}}{2p_{i+}}\bigg]\xi_i(p_i)\,.
\label{psi}
\end{equation}
The first term on the left-hand side of \eqref{eq:relperp} 
reproduces the $C^{A1\mu}_{i,\textrm{spinor}}(\{p_{i+}\})$ piece of 
\eqref{A1}, hence, to $\Lo(\lambda^2)$, \eqref{eq:relperp} 
gives
\begin{equation}
\bar\xi_i\,\frac{\p}{\p p_{\mu i\perp}}\A_i\bigg|_{p^{\mu}_{i\perp}=0} = \bar\xi_{1...n}\,C^{A1\mu}_{i,\textrm{coll.}}(\{p_{i+}\})\,,
\end{equation}
which proves \eqref{eq:perprel}.

Therefore the SCET result \eqref{softfromscet} is precisely the $i$-th summand of the soft theorem \eqref{softfer}. Summing over all $i$, we obtain the full expression. 
This shows that the colour-kinematics duality at the Lagrangian level indeed is the origin of the the colour-kinematics duality apparent in the LBK theorem and the gravitational next-to-soft theorem. 


\section{Conclusion and outlook}

We find it interesting that colour-kinematics duality, 
and resulting 
double-copy relations between gauge-theory and gravitational 
amplitudes, can be made manifest at the level of Lagrangians 
rather than amplitudes, once one fixes light-cone gauge, 
in which only the physical degrees of freedom appear 
as dynamical fields. Moreover, the duality can be 
constructed in a very direct way from the simple mapping 
\eqref{vector} of the colour to the diffeomorphism 
generators together with a pairing prescription for 
products of fields. 

The new prescription also holds for couplings to matter, 
and we exemplified this by first double-copying the scalar and fermionic QCD Lagrangians to obtain the Lagrangians for gravity coupled to scalar and Dirac fermion matter, respectively, to trilinear order. The mapping was finally applied to the soft-collinear 
effective QCD Lagrangian including subleading-power 
interactions to obtain the corresponding SCET gravity 
terms. These indeed produce the correct soft and 
next-to-soft theorems for the emission of gravitons, 
here for Dirac fermions.

The presented prescription is guided by the idea that gravity can be interpreted as gauge theory with the diffeomorphism gauge group. Indeed, we found that the Einstein-Hilbert Lagrangian up to trilinear order can be derived from the Yang-Mills Lagrangian in light-cone gauge. The known double-copy relation  \eqref{relation} between the three-point vertices then follows from this result. One of the reasons why the Lagrangian double copy is interesting is because it should expose the underlying mechanism which allows the amplitude double-copy to work in the first place. Our result can be considered as the starting point for such an explanation. Because the three-point amplitude is all that is needed to recursively construct any higher-point tree-level gravitational amplitude \cite{Benincasa:2007qj}, just as for tree-level Yang-Mills amplitudes \cite{BCFW1,BCFW2}, the trilinear Lagrangian double-copy provides a candidate for the conceptual underpinning of the BCJ double copy \cite{BCJ}, since one then has an iterative correspondence between gauge-theory and gravitational amplitudes.

It is a common feature of all previous realizations of colour-kinematics duality or double-copy structures at Lagrangian level \cite{BCJ,BRSTDC,Borsten:2021hua,Borsten:2021rmh,EHcubic} that they apply straightforwardly only to the three-point vertices. Since the Einstein-Hilbert Lagrangian contains vertices of arbitrary order, whereas the Yang-Mills Lagrangian does not, previous approaches to realize higher-point vertices at Lagrangian 
level point towards a non-local rearrangement as the starting point for applying a colour-kinematics duality mapping, as was done in \cite{BCJ} for the first time. In the spinor-helicity formalism the generalization to higher-point amplitudes is by-passed by their recursive construction starting from the three-point amplitude with complex momenta.

Although a treatment of four-point interactions is beyond 
the scope of this work, we note that the presented 
framework already contains three main ingredients. The first is the straightforward application of the mapping prescription to the quartic Yang-Mills Lagrangian in light-cone gauge \eqref{quadritr}, which generates quartic terms in the double-copy field $H$. The second is to take into account that the presented prescription has some non-linearity already built in. The application of the mapping to \eqref{a-}, 
\begin{equation}
\cA^a_-T^a=-\frac{2}{\p_+}\p^i\cA^a_iT^a-2ig_s\frac{1}{\p^2_+}[\cA^a_iT^a,\p_+\cA^{b,i}T^b]\,,
\end{equation}
which we rewrote here by multiplying \eqref{a-} with $T^a$, summing over $a$ and using \eqref{liea}, results in
\begin{eqnarray}
-iH_{-\mu}\p^{\mu}&=&-i\,\bigg\{-\frac{2}{\p_+}\p^iH_{i\mu}+\frac{\kappa}{2}\frac{1}{\p^2_+}\bigg[\frac{1}{2}H_{i-}\p^2_+H^i_{\;\mu}+H_{ij}\p^j\p_+H^i_{\;\mu}\nonumber\\
&&\hspace*{1cm}-\,\frac{1}{2}\p_+H_{i-}\p_+H^i_{\;\mu}-\p_+H_{ij}\p^jH^i_{\;\mu}\bigg]\bigg\}\,\p^{\mu}\,.
\end{eqnarray}
This yields a non-linear equation in $\kappa$ for $H_{-i}$, which needs to be solved perturbatively, generating an infinite number of terms in powers of the transverse components $H_{ij}$. Since we already used the leading-order solution of the above for $H_{-i}$,
\begin{equation}
H_{-i}=-\frac{2}{\p_+}\p^jH_{ji}+\Lo(\kappa)\,,
\end{equation}
to eliminate it in the construction of the trilinear double-copy Lagrangian \eqref{dualtri}, we will thus generate additional terms at higher order in $\kappa$ from the trilinear Lagrangian, and later also from the quartic one. Therefore, we see that this non-linearity naturally generates interaction terms at arbitrarily high order in $\kappa$, not just $\Lo(\kappa^2)$. The thus obtained terms then need to be written in terms of the graviton, Kalb-Ramond and dilaton fields by means of the decomposition \eqref{decH}. Third, the $\ch\ch\ch\ch$-terms generated by eliminating the $\ch\ch\phi$-interaction term using the dilaton equation-of-motion in section \ref{elimdilaton} need to be taken into account, which will contribute to the four-point gravity Lagrangian. After all this has been done the result needs to be compared to the quartic Einstein-Hilbert Lagrangian in light-cone gauge with redundant field components $\ch_{--}$, $\ch_{-i}$, $\ch$ integrated out. 
We leave the systematic computations and the comparison with higher than trilinear graviton vertices in light-cone gauge for future work.

\vspace*{0.5em}
\noindent
\subsubsection*{Acknowledgement}
This work was supported in part by the Excellence Cluster ORIGINS
funded by the Deutsche Forschungsgemeinschaft (DFG, German Research Foundation)
under Germany's Excellence Strategy - EXC-2094 - 390783311.


\begin{appendices}

\section{Full trilinear $\mathcal{N}=0$ supergravity Lagrangian}
\label{fullphystri}
The full trilinear Lagrangian in terms of the graviton, Kalb-Ramond and dilaton fields obtained by substituting the decomposition of the double-copy field \eqref{decH} into \eqref{dualtri} reads:
\begin{eqnarray}
\Ld^{(\kappa)}_{\textrm{grav}}&=&
\frac{\kappa}{2}\bigg[\ch_{ij}\p^2_+\ch^{ij}\frac{\p_k\p_l}{\p^2_+}\ch^{kl}-2\ch_{ij}\p_+\p^k\ch^{ij}\frac{\p^l}{\p_+}\ch_{kl}+\ch_{ij}\ch_{kl}\p^k\p^l\ch^{ij}-2\ch_{il}\ch_{kj}\p^k\p^l\ch^{ij}
\nonumber\\
&&-\,4\ch_{il}\p_+\ch^{ij}\frac{\p^k\p^l}{\p_+}\ch_{kj}+B_{ij}\p^2_+B^{ij}\frac{\p_k\p_l}{\p^2_+}\ch^{kl}-2B_{ij}\p_+\p^kB^{ij}\frac{\p^l}{\p_+}\ch_{kl}-2B_{il}B_{kj}\p^k\p^l\ch^{ij}\nonumber\\
&&+\,B_{ij}\ch_{kl}\p^k\p^lB^{ij}-2\ch_{il}B_{kj}\p^k\p^lB^{ij}-4B_{il}\p_+\ch^{ij}\frac{\p^k\p^l}{\p_+}B_{kj}-4\ch_{il}\p_+B^{ij}\frac{\p^k\p^l}{\p_+}B_{kj}\nonumber\\
&&-\,4B_{il}\p_+B^{ij}\frac{\p^k\p^l}{\p_+}\ch_{kj}+\frac{1}{2}B_{ij}\p^2_+B^{ij}\frac{\p_k\p^k}{\p^2_+}\phi-B_{ij}\p_+\p^kB^{ij}\frac{\p_k}{\p_+}\phi+\frac{1}{2}\phi B_{ij}\p_k\p^kB^{ij}\nonumber\\
&&-\,2B_{il}B_k^{\phantom{k}i}\p^k\p^l\phi-2B_{il}\p_+\phi\frac{\p^k\p^l}{\p_+}B_k^{\phantom{k}i}+\frac{1}{2}\ch_{ij}\p^2_+\ch^{ij}\frac{\p_k\p^k}{\p^2_+}\phi-\ch_{ij}\p_+\p^k\ch^{ij}\frac{\p_k}{\p_+}\phi+\frac{1}{2}\phi\ch_{ij}\p_k\p^k\ch^{ij}\nonumber\\
&&+\,\frac{1}{2}\phi\p^2_+\phi\frac{\p_k\p_l}{\p^2_+}\ch^{kl}-\phi\p_+\p^k\phi\frac{\p^l}{\p_+}\ch_{kl}+\frac{1}{2}\ch^{ij}\phi\p_i\p_j\phi+\frac{1}{4}\phi\p^2_+\phi\frac{\p_k\p^k}{\p^2_+}\phi-\frac{1}{2}\phi\p_+\p^k\phi\frac{\p_k}{\p_+}\phi\nonumber\\
&&+\,\frac{1}{4}\phi^2\p_k\p^k\phi\bigg]\,.
\end{eqnarray}
As stated in the main text, the $\ch\ch\phi$-terms are effectively $\Lo(\kappa^2)$.

\section{SCET QCD and effective Lagrangian at subleading power}
\label{DCSCET}

In the following, we present a brief introduction to the position-space description of soft-collinear effective theory.
For more details, we refer to the original works \cite{Beneke:2002ni, Beneke:2002ph}.

Soft-collinear effective theory is a framework that involves energetic (``collinear'') and soft particles.
The collinear particles are characterised by their large momentum along the light-like direction $n_-$.
We also introduce a second light-like reference vector $n_+$ such that
\begin{equation}
    n_{\pm}^2=0\,,\quad n_+\cdot n_- = 2\,.
\end{equation}
With these reference vectors, the momentum $p$ is decomposed as
\begin{equation}\label{eq::App::MomentumScaling}
    p^\mu = n_+\cdot p \frac{n_-^\mu}{2} + p_\perp^\mu + n_- \cdot p \frac{n_+^\mu}{2}\,.
\end{equation}
The individual components scale as
\begin{equation}
    n_+\cdot p \sim Q\,,\quad p_\perp\sim\lambda Q\,,\quad n_-\cdot p \sim\lambda^2 Q\,,
\end{equation}
where $Q$ is a hard scale (often set to 1, as we do in the following) and $\lambda\ll 1$.
Soft particles are characterised by homogeneous momenta $k_s\sim\lambda^2$.

In the effective theory, each full-theory field is split into a soft and a collinear mode for each existing collinear direction.
In position space formalism, the scaling \eqref{eq::App::MomentumScaling} applies to derivatives of collinear fields (and similarly for soft).
In QCD, the theory contains the collinear gluon components $n_+\cdot A_c$, $A_{c\perp}$ and $n_-\cdot A_c$, which scale as
\begin{equation}
    \left( n_+\cdot A_c, A_{c\perp}, n_-\cdot A_c \right) \sim (1,\lambda,\lambda^2)\,,
\end{equation}
and a soft gluon field $A_s\sim\lambda^2$.
In addition, we introduce collinear and soft quark fields $\psi_c$ and $q_s$.
The collinear quark field contains two large and two small components.
Using the projection operators
\begin{equation}
    \frac{\slashed n_- \slashed n_+}{4}\,,\quad \frac{\slashed n_+ \slashed n_-}{4}
\end{equation}
they are defined as
\begin{equation}
    \psi_c(x) = \xi(x) + \eta(x)\,,\quad \xi(x) \equiv \frac{\slashed n_- \slashed n_+}{4}\psi_c(x)\,,\quad \eta(x) \equiv \frac{\slashed n_+ \slashed n_-}{4}\psi_c(x)\,,
\end{equation}
where $\xi(x)\sim \lambda$ and $\eta(x)\sim\lambda^2$.
The soft quark scales as $q_s\sim\lambda^3$.

In addition, the effective theory makes use of various Wilson lines.
Relevant in the following is only the collinear Wilson line $W_c$, defined by
\begin{equation}\label{eq::app::Wc}
    W_c(x) \equiv P \exp \left( ig \int_{-\infty}^0 \der s\: n_+\cdot A_c(x+sn_+)\right)\,,
\end{equation}
where $P$ denotes path-ordering. This Wilson line is used to define manifestly collinear-gauge invariant building blocks and to eliminate the $n_+\cdot A_c$ component of the gluon field in the operator basis.

In position-space formalism, the argument $x$ of collinear fields scales as
\begin{equation}
    n_+\cdot x \sim \frac{1}{\lambda^2}\,,\quad x_\perp \sim \frac{1}{\lambda}\,,\quad n_-\cdot x \sim 1\,.
\end{equation}
For soft fields, the scaling is homogeneous as $x \sim\frac{1}{\lambda^2}$.
This implies that soft fields in soft-collinear interactions must be multipole expanded around $x_- \equiv n_+\cdot x \frac{n_-}{2}$.
This multipole expansion leads to explicit factors of $x$ in the subleading Lagrangian interactions.
The multipole-expanded Lagrangian can be split into the Yang-Mills part $\mathcal{L}_{\rm YM}$ and a part involving the quark matter fields $\mathcal{L}$ \cite{Beneke:2002ni}.
Relevant for us is only the latter one.
It is given as a power series in $\lambda$,
\begin{equation}
    \mathcal{L} = \overline{\xi}\left( in_-\cdot D + i \slashed D_\perp \frac{1}{i n_+\cdot D}i\slashed D_\perp\right) \frac{\slashed n_+}{2}\xi + \overline{q} i\slashed D_s q + \mathcal{L}^{(1)}_\xi + \mathcal{L}^{(2)}_\xi + \mathcal{L}^{(1)}_{\xi q} + \mathcal{L}^{(2)}_{\xi q}\,,
    \label{masterl}
\end{equation}
where
\begin{align}
    D^\mu &= \partial^\mu - ig_s A_c^\mu(x) - ig_s n_-\cdot A_s(x_-)\frac{n_+^\mu}{2}\,,\\
    D_s^\mu &= \partial^\mu - ig_s A_s^\mu(x)\,.
\end{align}
The terms $\mathcal{L}^{(n)}$ denote the $n$-th power subleading interactions, which contain only collinear quarks, denoted by $\mathcal{L}_\xi$, or both soft and collinear quarks $\mathcal{L}_{\xi q}$.
The terms $\mathcal{L}_\xi^{(n)}$ up to $\mathcal{O}(\lambda^2)$ are given by \cite{Beneke:2002ni}

\begin{flalign}
\Ld^{(1)}_{\xi}=&\bar\xi\Big(x^{\mu}_{\perp}n^{\nu}_-W_cg_sF_{s,\mu\nu}W^{\dagger}_c\Big)\frac{\slashed{n}_+}{2}\xi\,,\label{l1}\\
\Ld^{(2)}_{\xi}=&\frac{1}{2}\bar\xi\Big((n_-\cdot x)n^{\mu}_+n^{\nu}_-W_cg_sF_{s,\mu\nu}W^{\dagger}_c+x^{\mu}_{\perp}x_{\rho\perp}n^{\nu}_-W_c[D^{\rho}_s,g_sF_{s,\mu\nu}]W^{\dagger}_c\Big)\frac{\slashed{n}_+}{2}\xi\nonumber\\
&+\frac{1}{2}\bar\xi\bigg(i\slashed{D}_{c\perp}\frac{1}{in_+\cdot D_c}x^{\mu}_{\perp}\gamma^{\nu}_{\perp}W_cg_sF_{s,\mu\nu}W^{\dagger}_c+x^{\mu}_{\perp}\gamma^{\nu}_{\perp}W_cg_sF_{s,\mu\nu}W^{\dagger}_c\frac{1}{in_+\cdot D_c}i\slashed{D}_{c\perp}\bigg)\frac{\slashed{n}_+}{2}\xi\,.\label{l2}
\end{flalign}
Soft fields in the Lagrangian are always evaluated at $x_-$, possibly after taking derivatives.
This Lagrangian is not modified by radiative corrections \cite{Beneke:2002ph}.
Note that purely collinear interactions are all leading-power, while the subleading interactions are soft-collinear.
Due to the multipole expansion and required redefinitions \cite{Beneke:2002ni}, the soft gluon appears only as $n_-\cdot A_s(x_-)$ inside the soft-covariant derivative, or in terms of the field-strength tensor $F_{s,\mu\nu}$ in the subleading interactions.

\section{Feynman rules for a Dirac fermion coupled to gravity up to $\Lo(\lambda^2)$ in the CK-dual description}
\label{sec:gravitySCETFeynmanrule}

To obtain the interaction terms between fermions and gravitons from \eqref{l0dc}-\eqref{l2dc} we replace $S_{\mu\nu}\rightarrow s_{\mu\nu}$ in these terms and obtain the following symmetrized interaction vertex Feynman rules up to $\Lo(\lambda^2)$ between two collinear fermions and a soft graviton (fermion momenta flowing along the arrows, graviton momentum ingoing):
\vspace{0.2cm}
\begin{equation}
\raisebox{-1.87cm}{\includegraphics[width=0.3\textwidth]{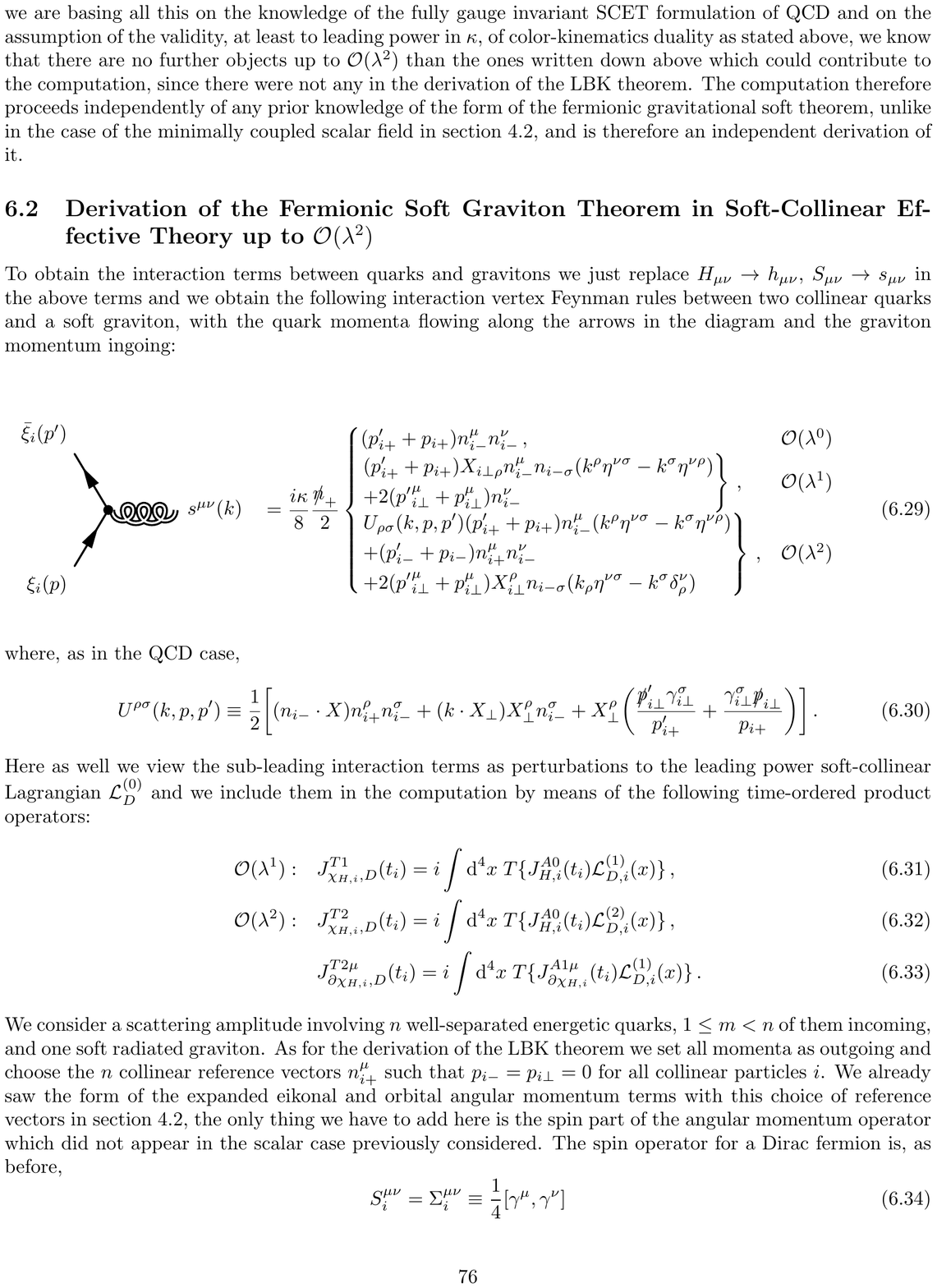}}=-\frac{i\kappa}{8}\frac{\slashed{n}_+}{2}\begin{cases}
	\displaystyle (p'_{i+}+p_{i+})n^{\mu}_{i-}n^{\nu}_{i-}\,, &\Lo(\lambda^0)\\[0.4cm]
	\begin{rcases}
	\hspace{-0.35cm}&\displaystyle (p'_{i+}+p_{i+})X_{i\perp\rho}n^{\mu}_{i-}n_{i-\sigma}(k^{\rho}\eta^{\nu\sigma}-k^{\sigma}\eta^{\nu\rho})\;\\
	\hspace{-0.35cm}&\displaystyle+2({p'}^{\mu}_{i\perp}+p^{\mu}_{i\perp})n^{\nu}_{i-}
	\end{rcases}\,, 	&\Lo(\lambda^1)\\[0.9cm]
	\begin{rcases}
	\hspace{-0.35cm}&\displaystyle U_{\rho\sigma}(k,p,p')(p'_{i+}+p_{i+})n^{\mu}_{i-}(k^{\rho}\eta^{\nu\sigma}-k^{\sigma}\eta^{\nu\rho})\;\\
	\hspace{-0.35cm}&\displaystyle+(p'_{i-}+p_{i-})n^{\mu}_{i+}n^{\nu}_{i-}\\
	\hspace{-0.35cm}&\displaystyle+2({p'}^{\mu}_{i\perp}+p^{\mu}_{i\perp})X^{\rho}_{i\perp}n_{i-\sigma}(k_{\rho}\eta^{\nu\sigma}-k^{\sigma}\delta^{\nu}_{\rho})
	\end{rcases}\,, 		&\Lo(\lambda^2)
	\end{cases}
\label{softcollvertex}
\end{equation}
\vspace{0.2cm}
\noindent where
\begin{equation}
U^{\rho\sigma}(k,p,p')\equiv\frac{1}{2}\bigg[(n_{i-}\cdot X)n^{\rho}_{i+}n^{\sigma}_{i-}+(k\cdot X_{\perp})X^{\rho}_{\perp}n^{\sigma}_{i-}+X^{\rho}_{\perp}\bigg(\frac{\slashed{p}'_{i\perp}\gamma^{\sigma}_{i\perp}}{p'_{i+}}+\frac{\gamma^{\sigma}_{i\perp}\slashed{p}_{i\perp}}{p_{i+}}\bigg)\bigg]\,.
\end{equation}
In the computation in Section \ref{comp} all momenta are outgoing, hence 
\begin{equation}
X^{\mu}_i\equiv-\frac{\p}{\p p_{i\mu}}(2\pi)^4\delta^{(4)}\bigg(\sum_{j=1}^n p_j\bigg)\,,
\end{equation}
and each $x^{\mu}$ in the Lagrangian turns into a factor $iX^{\mu}$ in the Feynman rule. The momentum space collinear quark Feynman propagator results from the quark kinetic term in \eqref{masterl}, which is unchanged by the mapping, and reads
\begin{equation}
\Delta_F(k)=\frac{in_+\cdot k}{k^2+i\epsilon}\frac{\slashed{n}_-}{2}\,.
\end{equation}

\end{appendices}

\bibliographystyle{JHEP}
\bibliography{paperbib}{}

\end{document}